\documentclass[fleqn,usenatbib]{mnras}

\usepackage[T1]{fontenc}

\DeclareRobustCommand{\VAN}[3]{#2}
\let\VANthebibliography\thebibliography
\def\thebibliography{\DeclareRobustCommand{\VAN}[3]{##3}\VANthebibliography}

\usepackage{graphicx}	
\usepackage{amsmath}	
\usepackage{amssymb}	
\usepackage{xcolor}
\usepackage{gensymb}
\usepackage[skip=0.5ex]{subcaption}
\usepackage{newtxtext,newtxmath}
\definecolor{mygreen}{rgb}{0,0.4,0}

\title[Massive YSOs in M\,33]{Massive young stellar objects in the Local Group spiral galaxy M\,33 identified using machine learning}

\author[Kinson et al.]{
David A. Kinson,$^{1}$\thanks{Email: d.a.kinson@keele.ac.uk}
Joana M. Oliveira,$^{1}$
and Jacco Th. van Loon$^{1}$
\\
$^{1}$ Lennard-Jones Laboratories, School of Chemical and Physical Sciences, Keele University, ST5 5BG, UK
}

\date{Accepted XXX. Received YYY; in original form ZZZ}

\pubyear{2022}

\begin{document}
\label{firstpage}
\pagerange{\pageref{firstpage}--\pageref{lastpage}}
\maketitle

\begin{abstract}
We present a supervised machine learning classification of stellar populations in the Local Group spiral galaxy M\,33. The Probabilistic Random Forest (PRF) methodology, previously applied to populations in NGC\,6822, utilises both near and far-IR classification features. It classifies sources into nine target classes: young stellar objects (YSOs), oxygen- and carbon-rich asymptotic giant branch stars, red giant branch and red super-giant stars, active galactic nuclei, blue stars (e.g. O-, B- and A-type main sequence stars)
, Wolf-Rayet stars and Galactic foreground stars. Across 100 classification runs the PRF classified 162,746 sources with an average estimated accuracy of $\sim$\,86\,per\,cent, based on confusion matrices. 
We identified 4985 YSOs across the disk of M\,33, applying a density-based clustering analysis to identify 68 star forming regions (SFRs) primarily in the galaxy's spiral arms. SFR counterparts to known H\,{\sc ii} regions were recovered, with $\sim$\,91\,per\,cent of SFRs spatially coincident with giant molecular clouds identified in the literature. Using photometric measurements, as well as SFRs in NGC\,6822 with an established evolutionary sequence as a benchmark, we employed a novel approach combining 
ratios of [H$\alpha$]$/$[24$\mu$m] and [250$\mu$m]$/$[500$\mu$m] to estimate the relative evolutionary status of all M\,33 SFRs. Masses were estimated for each YSO ranging from 6\,$-$\,27\,M$_\odot$. Using these masses, we estimate star formation rates based on direct YSO counts of 0.63\,M$_\odot$\,yr$^{-1}$ in M\,33's SFRs, 0.79\,$\pm$\,0.16\,M$_\odot$\,yr$^{-1}$ in its centre and 1.42\,$\pm$\,0.16\,M$_\odot$\,yr$^{-1}$ globally. 
\end{abstract}




\section{Introduction}

Studies of the galaxy M\,33 and its stellar populations began with \citet{1926ApJ....63..236H} yet nearly 100 years hence a comprehensive study of resolved star formation across the galaxy is still unavailable. M\,33 is the third largest galaxy in the Local Group ($M_{\rm gas}$\,$\sim$\,3\,$\times$\,10$^{9}$\,M$_\odot$, \citealt{2003MNRAS.342..199C}; $M_{*}$\,$\sim$\,5.5\,$\times$\,10$^9$\,M$_\odot$, \citealt{2014A&A...572A..23C,2017AJ....154...41K}), after the Milky Way and M\,31. M\,33 lies at a distance of $\sim$\,850\,kpc ($\mu{_{_{\rm M33}}}$\,=\,24.67\,mag, \citealt{2014AJ....148...17D}) and extends to an apparent size of approximately 60\arcmin\,$\times$\,35\arcmin \citep{2003A&A...412...45P}. Its relatively face-on inclination ($i$\,=\,54\degree,\, \citealt{1991rc3..book.....D}) makes M\,33 a more favourable target to study the entirety of a spiral galaxy's disk over the larger and similarly distant M\,31 which is seen nearly edge on \citep[e.g.][]{2001ChPhL..18.1420M}. 

The metallicity of M\,33 is around half-solar \citep[e.g.][]{2018A&A...612A..51B}, similar to that of the LMC \citep[see figure\,1 of][]{2021ApJS..253...53W}. The metallicity of M\,33 varies across the disk, with a negative gradient with increasing galactocentric radius well documented \citep[e.g.][]{1971ApJ...168..327S,2009A&A...506.1137C,2010A&A...512A..63M,2022ApJ...925...76A}; however its steepness is debated, with recent results favouring a shallower slope \citep{2022ApJ...925...76A}. A negative gradient supports an inside-out model of disk formation \citep{2009A&A...506.1137C,2009ApJ...695L..15W}, supported in M\,33's by the observed star formation history radial profiles \citep{2009ApJ...695L..15W,2017MNRAS.464.2103J}. The radial stellar age profile has been reported to reverse at radii larger than 9\,kpc beyond the break in optical brightness of the disk \citep{2009ApJ...695L..15W,2011MNRAS.410..504B,2018MNRAS.480.4455M}. A similar break in the gas velocity profiles is observed \citep[e.g.][]{2014A&A...572A..23C,2015MNRAS.449.4048K}, however a link between these has not been definitively made.

Whilst the outer gas distribution of M\,33 is warped \citep{1976ApJ...204..703R,2014A&A...572A..23C}, likely by a previous minor interaction with M\,31 \citep{2018ApJ...864...34S}, the disk within 9\,kpc appears relatively undisturbed \citep{2022AJ....163..166Q}. M\,33 is a flocculent spiral, with two primary spiral arms plus four additional fragmentary arms either side of the centre branching from, and filling in between, the primary arms \citep{1980ApJS...44..319H}. M\,33 is generally not categorised as a barred galaxy, however recent observations suggest the presence of a weak bar within the bright central region \citep{2021ApJS..253...53W,2022arXiv220611393L}. Whilst there is no strong central bulge in M\,33 \citep[e.g.][]{1991PASP..103..609V} a nuclear star cluster is present, with star formation thought to have occurred there inside the last 40\,Myrs \citep{2002ApJ...569..204L,2011MNRAS.414.3394J}. 
The spiral arms of M\,33 can be traced in the distributions of H\,{\sc i} \citep{2010A&A...522A...3G} and CO \citep{2014A&A...567A.118D,2018A&A...612A..51B} emission, giant molecular clouds \citep[GMCs,][]{2017A&A...601A.146C} and bright young clusters \citep{1980ApJS...44..319H,2021ApJS..253...53W}. The multiple arms in flocculent galaxies have been suggested to support the model of dynamic spiral formation \citep{2014PASA...31...35D} over the quasi-static model \citep{1964ApJ...140..646L}. GMCs studied in M\,33 however show an evolutionary progression which is associated with quasi-static arm models \citep{2017A&A...601A.146C}, as gas accumulates at the potential minimum triggering cloud collapse \citep{1964ApJ...140..646L}. 

The arm structure is also well traced by the distribution of H\,{\sc ii} regions \citep{1980ApJS...44..319H,2022ApJ...925...76A}. M\,33 contains many prominent H\,{\sc ii} regions which have have been studied widely across M\,33 alongside GMCs \citep{2010A&A...522A...3G,2012ApJ...761...37M,2017A&A...601A.146C,2022ApJ...925...76A}. Resolved IR observations of ongoing star formation, i.e. of massive YSOs in M\,33 however have not been extended beyond NGC\,604 \citep[e.g.][]{2012AJ....143...43F}. NGC\,604 is the second most luminous H\,{\sc ii} region in the Local Group behind only 30\,Dor in the LMC \citep{2009ApJ...699.1125R,2012ApJ...761....3M}. Star formation in NGC\,604 has been well studied at many wavelengths \citep[e.g.][]{1999ApJ...514..188C,2007A&A...466..509T} including both near-IR studies of individual massive young stellar objects (YSOs) \citep{2012AJ....143...43F} and integrated mid-IR properties \citep{2009ApJ...699.1125R,2012ApJ...761....3M}. Triggered star formation events have been theorised in NGC\,604 \citep{2007A&A...466..509T,2018PASJ...70S..52T}, possibly driven by feedback from a population of around 200 O-type stars \citep{1996ApJ...456..174H}.

Machine learning offers a method by which sources in large, multi-dimensional data sets can be accurately classified. In the Local Group dwarf-irregular galaxy NGC\,6822, sites of ongoing star formation were identified from wide-scale near-IR survey data using probabilistic random forest (PRF) analysis \citep{2021MNRAS.507.5106K}. A combination of near-IR and far-IR classification features were used in NGC\,6822 to separate point sources into multiple object classes. The classifier achieved high levels of estimated accuracy ($\sim$\,90\,per\,cent) across all classes, with that of massive YSOs exceeding this \citep{2021MNRAS.507.5106K}. The existence of similar near-IR data covering the M\,33 disk \citep{2015MNRAS.447.3973J} offers the opportunity to extend the detailed analysis of ongoing star formation, for the first time, across the entire disk of a spiral galaxy.

The paper is organised as follows. Section\,\ref{sec:Data} introduces the archival data used in this work, Section\,\ref{sec:PRF} contains details of our PRF classification method. The results are presented in Section\,\ref{sec:Results}, in which the spatial distributions of the different source classes are described and star formation regions (SFRs) are identified. In Section\,\ref{sec:Disc} we discuss the properties of YSOs and SFRs identified in our analysis, in the context of the galaxy's structure. Finally, in Section\,\ref{sec:Conc} we summarise our findings.

\begin{figure*}
    \centering
    \includegraphics[width=16.5cm]{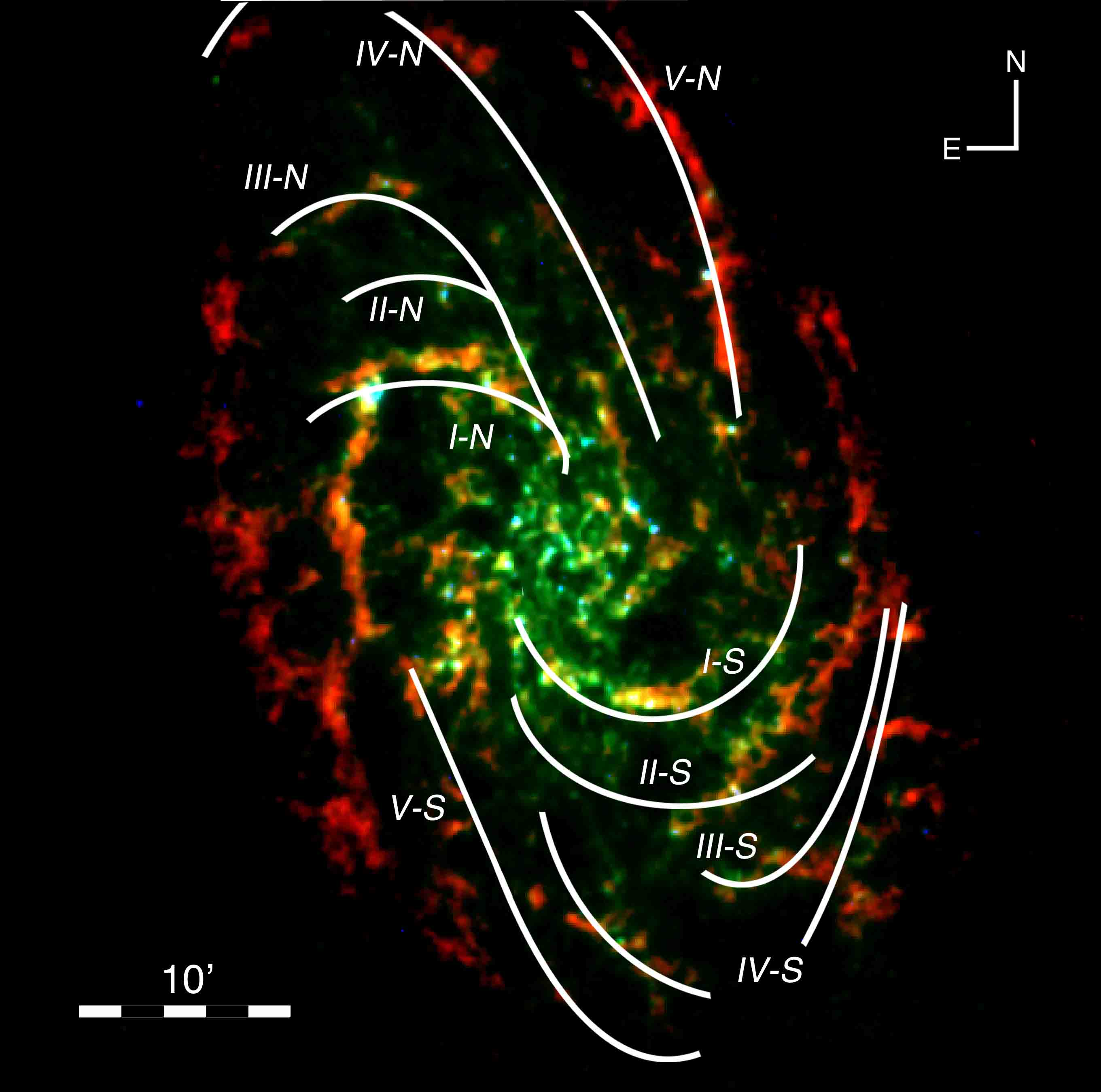}
    \caption{An RGB image of M\,33, showing VLA H\,{\sc i} \citep[red,][]{2010A&A...522A...3G}, 250\,$\mu$m {\it Herschel}-SPIRE \citep[green,][]{2010A&A...518L..67K}, 24\,$\mu$m {\it Spitzer}-MIPS \citep[blue,][]{2004AAS...204.3311E}. The figure covers the same footprint as the near-IR WFCAM catalogue of \citet{2015MNRAS.447.3973J}. The spiral arm identifications, adapted from \citet{1980ApJS...44..319H}, are shown in white.}
    \label{fig:GalView}
\end{figure*}

\section{Data}
\label{sec:Data}

The description of the data used in our analysis is divided in two parts: catalogues and images used for the PRF object classification (Sect.\,\ref{sec:PRF}), and images used for the subsequent analysis of star forming regions across the disk of M\,33 (Sect.\,\ref{ssec:M33Structure}).

\subsection{Data for object classification}

\subsubsection{Near-IR images and point-source catalogue}
\label{ssec:NIRData}

\begin{figure}
    \centering
    \includegraphics[width=\columnwidth]{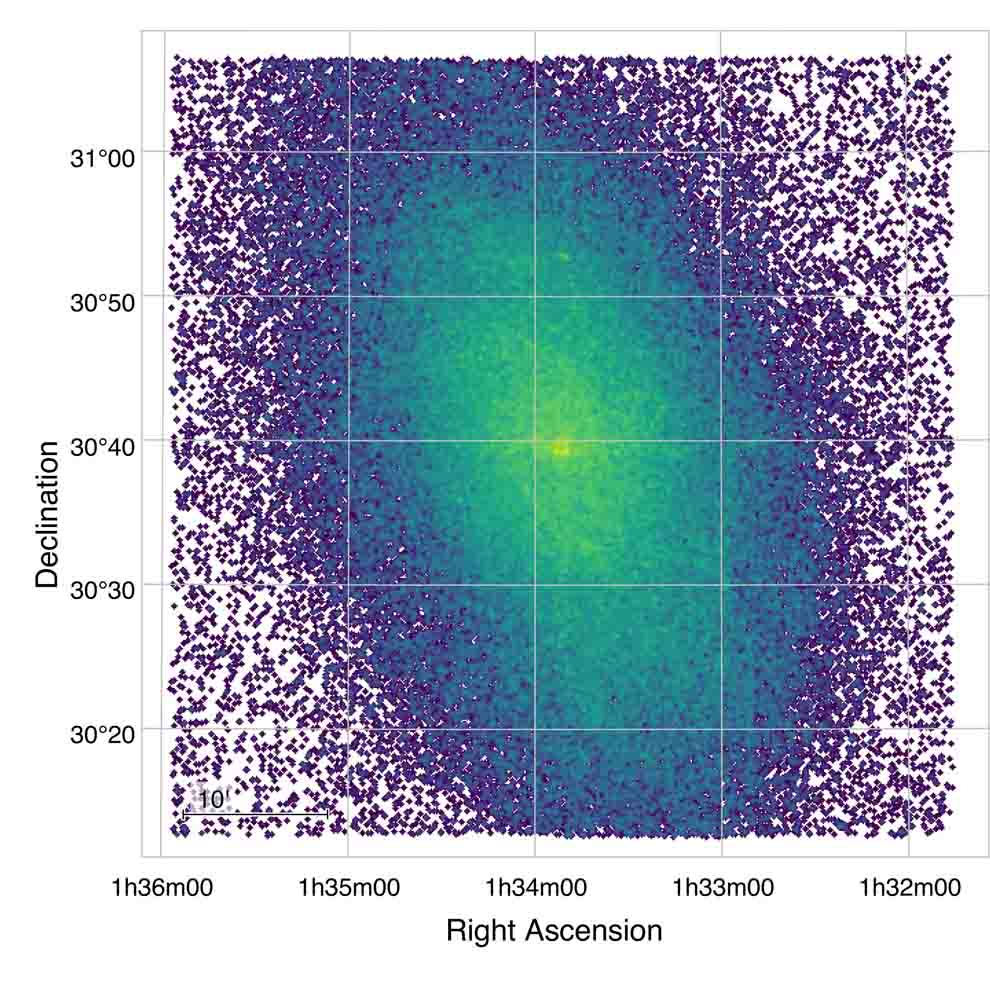}
    \includegraphics[width=\columnwidth]{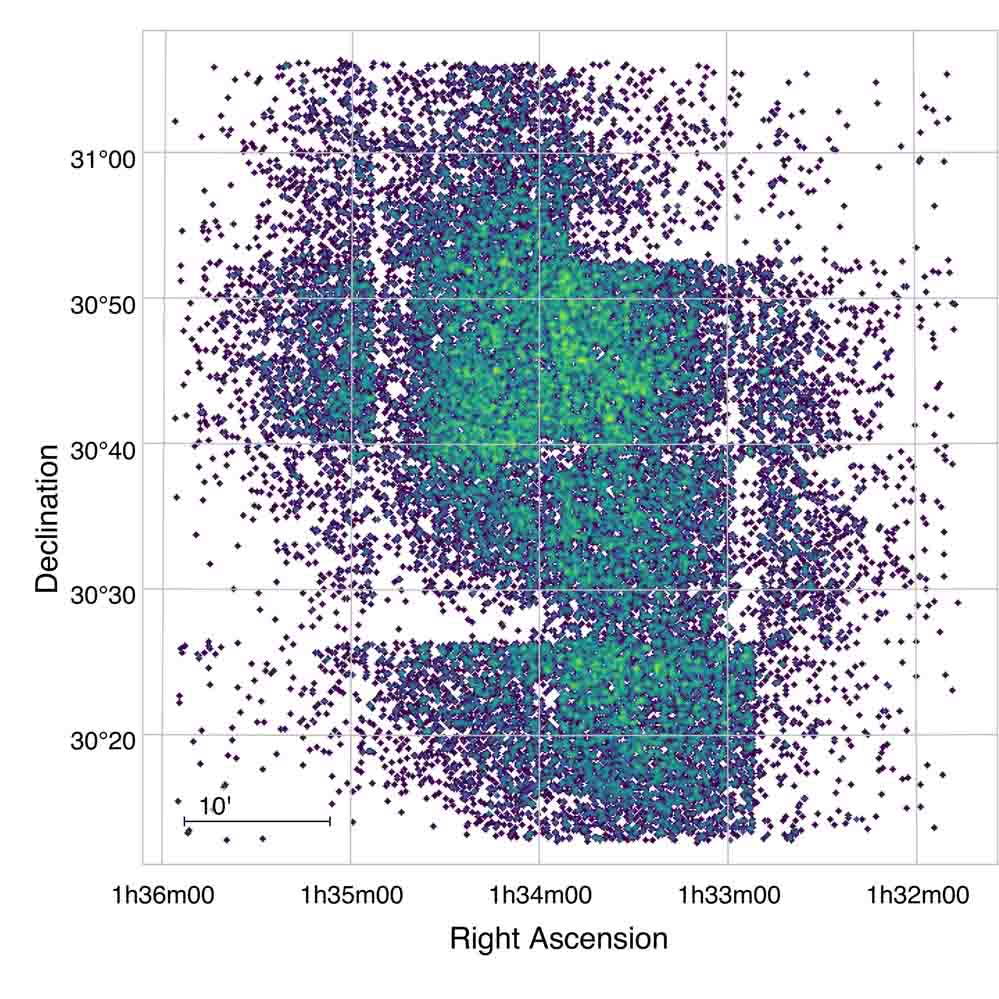}
    \caption{Hess diagrams of source density, brighter (top) and fainter (bottom) than $K_s$\,$=$\,19.2\,mag. The effect of variable depth in the catalogue across the field-of-view is clear at fainter magnitudes.}
    \label{fig:CatHess}
\end{figure}

\begin{figure}
    \centering
    \includegraphics[width=\columnwidth]{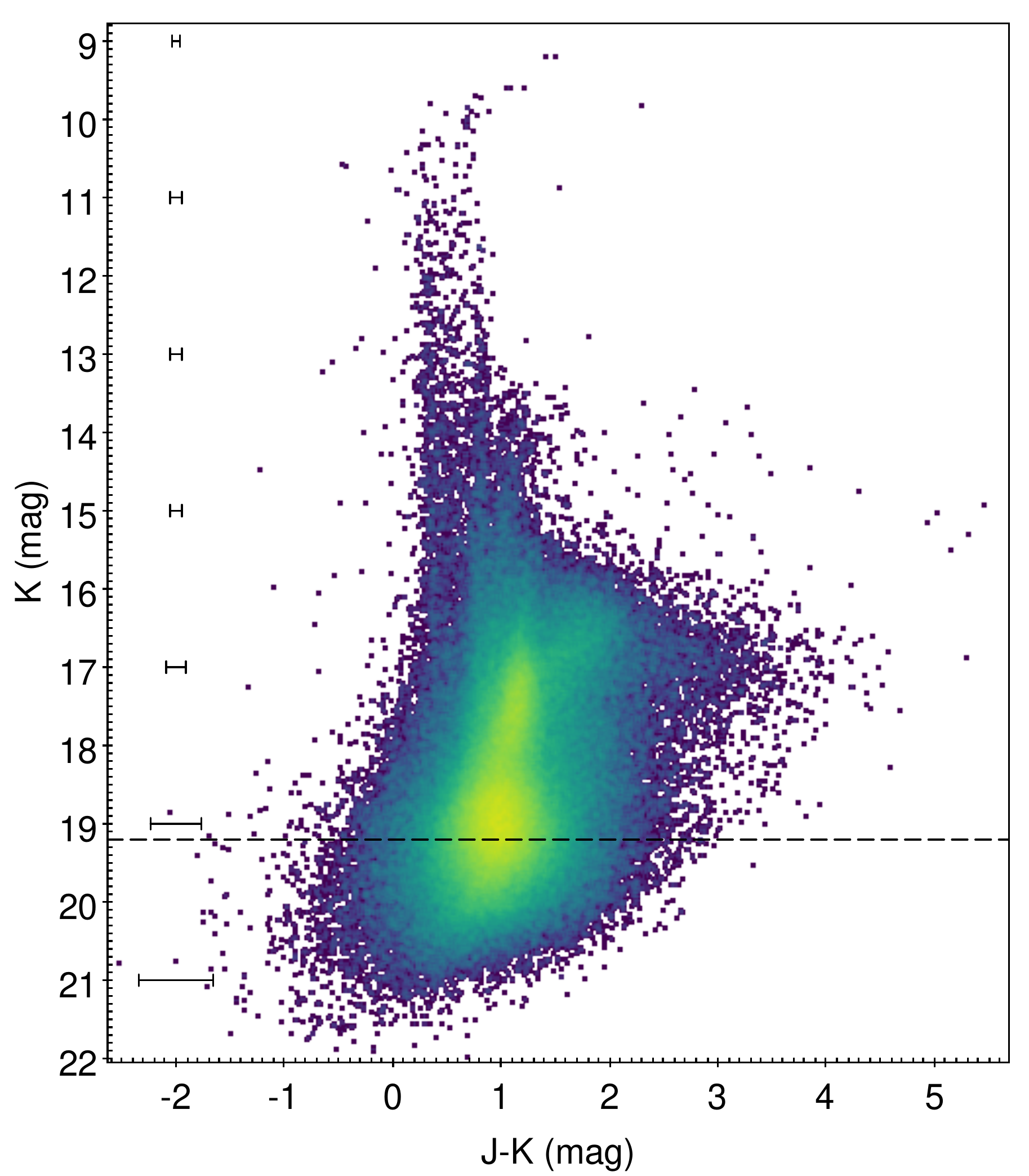}
    \caption{The near-IR catalogue presented in a CMD Hess diagram. Average error bars are shown. The dashed line at $K_s$\,$=$\,19.2\,mag indicates the magnitude at which the catalogue depth becomes very patchy (Fig.\,\ref{fig:CatHess}).}
    \label{fig:CatCMD}
\end{figure}

The near-IR catalogue for M\,33 was constructed by \citet{2015MNRAS.447.3973J} using data obtained on the United Kingdom Infrared Telescope (UKIRT) using the Wide Field Camera \citep[WFCAM,][]{2007A&A...467..777C}. Four separate pointing observations were obtained to cover a $\sim$\,0.89\,deg$^2$ sky area ($\sim$\,13\,kpc\,$\times$\,13\,kpc) at a resolution of 0.4\,arcsec per pixel. Multi-epoch observations were made as part of a monitoring programme over dates from September 2005 to October 2007. More details on the data reduction can be found in \citet{2015MNRAS.447.3973J}. They retrieved the photometric catalogues for each individual tile and epoch from the public WFCAM Science Archive (WSA)\,\footnote{\url{http://wsa.roe.ac.uk/}} and performed absolute and relative photometric calibration. In our analysis we make use of their catalogue of mean magnitudes of point sources towards M\,33 for source classification (see Sect.\,\ref{sssec:ClassOut}). The catalogue contains $\sim$\,245,000 sources. We set the additional requirement that a source must be detected in all three $JHK_s$-bands, reducing the number of near-IR sources to $\sim$\,163,000 sources. The $JHK_s$ 5$\sigma$ limiting magnitudes of the catalogue are 21.5, 20.6 and 20.5\,mag respectively. Source density of the catalogue is shown in Fig.\,\ref{fig:CatHess} and basic photometric properties are shown in a Colour-Magnitude Diagram (CMD) in Fig.\,\ref{fig:CatCMD}. 

Due to the construction of the catalogue, with data taken over multiple epochs and detector pointings, different regions of the science field-of-view reach varying depths. As shown in Fig.\,\ref{fig:CatHess}, the catalogue is uniform to depths of $K_s$\,$=$\,19.2 mag, beyond which the varying depth between detectors becomes apparent. Whilst these artefacts in the catalogue construction will not affect the accuracy of classification for individual sources it is important to note when analysing the spatial distribution of sources (see Sect.\,\ref{sssec:SpDist}).

\subsubsection{Far-IR images and measurements}
\label{ssec:FIRData}

Light emitted by hot young stars at UV wavelengths is reprocessed by surrounding dust and re-emitted at far-IR wavelength \citep[e.g.][]{2012AJ....143...74B}. To provide additional environmental information for each source in the near-IR catalogue we use the neighbourhood far-IR brightness as an indicator of proximity to star-formation activity \citep{2021MNRAS.507.5106K}. To this end, we used 70 and 160\,$\mu$m images obtained with the Photodetector Array Camera \& Spectrometer \citep[PACS,][]{2010A&A...518L...2P} onboard the {\it ESA Herschel Space Observatory} \citep[{\it Herschel},][]{2010A&A...518L...1P}, obtained as part of the {\it HERschel M\,33 Extended Survey} \citep[HERM33ES,][]{2010A&A...518L..67K}. The images were retrieved from the ESA {\it Herschel} Science Archive \footnote{\url{http://archives.esac.esa.int/hsa/whsa/}}. 

Point sources located both in NGC\,6822 and the Magellanic Clouds (MC) were used to train the PRF classifier (see Sect.\,\ref{ssec:Training Set} for full details). For NGC\,6822, the 70 and 160\,$\mu$m images \citep{2010A&A...518L..55G} were also retrieved from the {\it Herschel} Science Archive, as were the Magellanic Clouds 160\,$\mu$m images \citep{2013AJ....146...62M}. The Magellanic 70\,$\mu$m images \citep{2006AJ....132.2268M,2011AJ....142..102G} were obtained using the Multiband Imaging Photometer for Spitzer \citep[MIPS,][]{2004ApJS..154...25R} onboard the {\it Spitzer Space Telescope} ({\it Spitzer}, \citealt{2004ApJS..154....1W}), retrieved from the {\it Spitzer} Heritage Archive \footnote{\url{https://sha.ipac.caltech.edu/applications/Spitzer/SHA/}}. Small non-astrophysical bias levels in some of the Magellanic images were corrected for as described in \citet{2021MNRAS.507.5106K}.

At the position of each $K_s$-band source, an aperture of 30\,parsec radius (7.2\,arcsec for M\,33) was used to measure an average brightness. Photometry was performed using the {\sc PhotUtils} package for {\sc Python} \citep{larry_bradley_2020_4044744}. The size of this aperture is the same as used in NGC\,6822 \citep{2021MNRAS.507.5106K} and was chosen based on the scale of emission in the far-IR images and typical molecular cloud scales \citep[e.g.][]{2014prpl.conf..149T}.

\subsection{Ancillary data}
\label{ssec:AncData}

Archival H$\alpha$, 24\,$\mu$m {\it Spitzer}-MIPS, and 250/500$\mu$m {\it Herschel}-Spectral and Photometric Imaging Receiver (SPIRE, \citealt{2010A&A...518L...3G}) images are used in our analysis to provide evolutionary information on the star forming regions, as discussed in Sect.\,\ref{ssec:M33Structure}.
The H$\alpha$ images of both M\,33 and NGC\,6822, retrieved from the NASA/IPAC Extragalactic Database (NED) \footnote{\url{https://ned.ipac.caltech.edu}}, were taken as part of a survey of Local Group galaxies \citep{2006AAS...209.2701M}; as described in \citet{2007AJ....133.2393M} the images were reduced and calibrated in a similar way and are therefore directly comparable with one another (see their tables\,1\ and 2). The {\it Spitzer}-MIPS 24\,$\mu$m mosaic images of both galaxies were retrieved from the {\it Spitzer} Heritage Archive (NGC\,6822: \citealt{2003PASP..115..928K}; M\,33: \citealt{2004AAS...204.3311E}). The {\it Herschel} Science Archive provided the 250/500$\,\mu$m SPIRE images, originally described in \citet{2010A&A...518L..67K} for M\,33 and \citet{2010A&A...518L..55G} for NGC\,6822.

\section{Probabilistic random forest (PRF)}
\label{sec:PRF}

A random forest classifier (RFC) is a robust and established tool for classification problems \citep[][]{breiman2001random}. We use an adaptation of the RFC developed by \citet{2019AJ....157...16R} called a probabilistic random forest (PRF). The PRF classifier improves on the RFC by taking into account feature uncertainties as well as allowing for the classification of sources with missing data. This both increases the accuracy of the classifier and the number of sources that can be classified \citep[][]{2019AJ....157...16R}. A more in-depth discussion of the difference in the methodologies for RFC and PRF classifiers is presented in \citet{2021MNRAS.507.5106K}; we follow their methodology that is summarised below. 

To classify the sources a set of six features were used: the near-IR $K_s$-band magnitude, three near-IR colours ($J$\,$-$\,$H$, $H$\,$-$\,$K_s$ and $J$\,$-$\,$K_s$) and two far-IR brightnesses at 70 and 160\,$\mu$m. To classify sources the PRF requires a set of sources of known type on which the algorithm is trained. These training set sources are then randomly split into training and testing samples, allowing for an estimate of the classifier's accuracy (see Sect.\,\ref{sssec:ConfMat}). Splitting is done on a 75\,per\,cent training, 25\,per\,cent test basis with the splitting applied globally to the training set rather than per each individual class. This random splitting can lead to some stochastic effects in the training data selection; these are mitigated by repeating the splitting over many runs with different random seeds. Where one class in the training set is disproportionately large, such that it dominates the randomly selected training sample, the accuracy of the classifier is negatively affected. We took steps to counteract this effect as described in Section\,\ref{ssec:TrainDown}. The following section details the sources selected for PRF training.

\subsection{Sources in the training set}
\label{ssec:Training Set}
The training set for the PRF consists of sources from nine target classes. These are Galactic foreground stars (FG), blue stars and yellow supergiant stars (BS), red supergiant stars (RSG), oxygen and carbon rich asymptotic giant branch stars (OAGB and CAGB), red giant branch stars (RGB), Wolf-Rayet stars (WR), massive young stellar objects (YSOs) and finally unresolved background galaxies (AGN). Other classes of objects are present in the M\,33 stellar population, but they are either dissimilar enough from YSOs that their misclassification will not contaminate the YSO sample or are rare (e.g. planetary nebulae) and so will not significantly impact the purity of the classified YSO sample. In the case of planetary nebulae, a PRF classification in NGC\,6822 misclassifies the few examples as AGN \citep[see section.\,3.4.7 of][]{2021MNRAS.507.5106K}. Including classes of rare objects would however adversely affect the accuracy of the PRF classifier (see Sect.\,\ref{ssec:TrainDown}).

The detailed selection criteria for each class is given in the following subsections. To maintain the purity of the training set stringent selection criteria are set, requiring sources identified in the literature to have been classified on the basis of methods other than broad-band photometry, e.g. spectroscopy, narrow-band indices or {\it Gaia} proper motions. In most instances, however, the catalogues from which training set sources are drawn do not completely cover the area of the near-IR catalogue. The M\,33 sources in the training sample were crossmatched to the near-IR catalogue using a radius of 0.5\,arcsec.

Most classes include exclusively sources in M\,33, with the exception of the AGN and RGB, that also include sources behind the MCs and in NGC\,6822. Training set YSOs come exclusively from the MCs and NGC\,6822. The near-IR data for NGC\,6822 and M\,33 are however comparable. We therefore believe that while the near-IR catalogue to be classified may be affected by source blending, such effects are on the whole also present in the training set data, providing the PRF with effective examples on which to learn.

\subsubsection{Foreground Galactic sources}
\label{sssec:TrainFG}

The training set of Galactic foreground contaminants includes sources from \citet{2016AJ....152...62M} with optical spectra consistent with Galactic dwarfs. They separate foreground dwarfs from B-, A-, F- and G-type supergiants by the shape and strength of their Balmer series lines, and the differing strengths of metallic lines (Si, Ca, K, Ti, Mg and Sr). Additionally we include near-IR sources with a {\it Gaia} EDR3\footnote{\url{https://www.cosmos.esa.int/web/gaia/earlydr3}} \citep{2020arXiv201201533G} counterpart if their proper motion is greater than 0.5\,mas\,yr$^{-1}$ in both RA and Dec components. Near-IR colour cuts at 0.3\,<\,$J$\,$-$\,$K_s$\,<\,0.9\,mag, defined using {\sc TRILEGAL} foreground simulations \citep{2005A&A...436..895G} towards M\,33, are then applied to remove spurious chance matches between the {\it Gaia} and near-IR catalogues. Whilst Galactic sources may be found outside these cuts, to ensure purity of the FG training set we select only sources in the conspicuous vertical foreground sequences (see Fig.\,\ref{fig:FG_DS_CMD}).

Foreground sources identified spectroscopically or with {\it Gaia} proper motions extend only to $K_s$\,$\sim$\,16.5\,mag; in order to accurately train the PRF however, foreground sources at magnitudes down to the limit of our near-IR catalogue ($K_s$\,$\sim$\,20.5\,mag) are needed. For this purpose, we used the foreground population simulated with {\sc TRILEGAL} already mentioned. The simulated foreground source magnitudes were perturbed in $J$-,$H$- and $K_s$ by an amount consistent with the average error bar in the near-IR catalogue at similar magnitudes. Foreground stars have no preferential location in the field of view, therefore, to generate far-IR measurements for these sources, apertures were placed randomly in the far-IR images and measurements taken as described in Section\,\ref{ssec:FIRData}. 

\subsubsection{Active galaxies}
\label{sssec:TrainAGN}

AGN have been shown to be significant contaminants in near-IR YSO samples due to their colour similarities \citep[e.g.][]{2013ApJ...778...15S, 2017MNRAS.470.3250J}. The strength of the far-IR emission as a measure of the proximity to star formation activity can help differentiate YSOs from contaminants such as AGN, as shown by \citet{2021MNRAS.507.5106K}. 
We start from the AGN training sample from \citet{2021MNRAS.507.5106K}, which is comprised of 89 background galaxies behind the SMC. This sample is classified using a variety of data across multiple wavelengths including X-Ray, UV, near-IR and radio \citep{2021IAUS..356..335P}. This AGN sample was augmented with 36 sources behind M\,33 taken from the latest update of the MILLIQUAS compilation \citep[the Million Quasars Catalog, version 7.2,][]{2021arXiv210512985F}.

\subsubsection{Asymptotic giant branch stars}
\label{sssec:TrainAGB}

\begin{figure}
    \centering
    \includegraphics[width=\columnwidth]{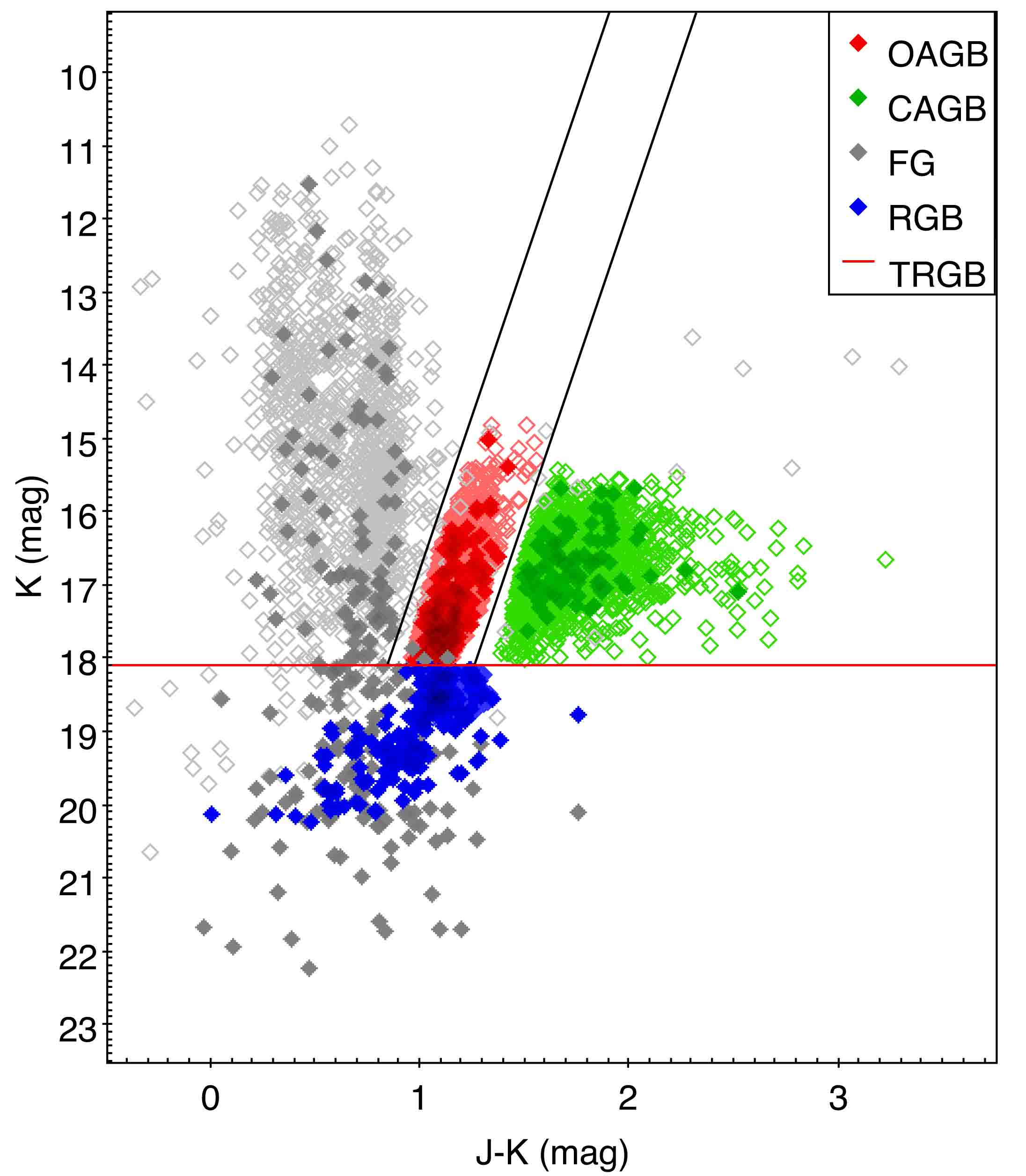}
    \caption{A CMD showing the four large classes which were down-sampled with the full set of data shown by open symbols and the down-sampled data by filled symbols. The parameter space for each class is well represented by the down-sampled data. The TRGB magnitude ($K_s$\,=\,18.11\,mag) and AGB colour-cuts adapted from \citet{2021ApJ...907...18R} are shown by the red and black lines respectively.}
    \label{fig:FG_DS_CMD}
\end{figure}

Asymptotic giant branch stars (AGBs) can display near-IR colours and magnitudes similar to bright massive YSOs. OAGBs and CAGBs have distinct magnitude and colour properties due to the composition of their circumstellar dust envelopes (see Fig.\,\ref{fig:FG_DS_CMD}) and thus are classified independently.

The AGB sample is based on the catalogue of {\it{V}} and {\it{I}} broadband and TiO and CN narrowband photometry towards M\,33 \citep{2005AJ....129..729R}. Using $V$\,$-$\,$I$ and CN\,$-$\,TiO colour cuts defined by \citet{2005AJ....129..729R} we identified both OAGBs and CAGBs from their catalogue. Both classes of AGB have $V$\,$-$\,$I$\,>\,1.8\,mag, with OAGBs having colours of CN\,$-$\,TiO\,<\,$-$0.2\,mag and CAGBs CN\,$-$\,TiO\,>\,0.3\,mag. From this sample we remove any sources with \textit{Gaia} proper motions consistent with a Galactic foreground dwarf (see Sect.\ref{sssec:TrainFG}). Sources with any spectroscopic classification of another type from \citet{2016AJ....152...62M} are also removed.

\citet{2021ApJ...907...18R} define near-IR colour and magnitude boundaries for both OAGB and CAGB sources in M\,33 (see their figure 10) which we adopted to refine the samples from \citet{2005AJ....129..729R}. These cuts are shown in Fig.\,\ref{fig:FG_DS_CMD}. We also select only sources brighter than the M\,33 tip of the RGB (TRGB) magnitude, $K_s$\,=\,18.11\,mag \citep{2021ApJ...907...18R}. Finally for the OAGBs we apply an upper magnitude limit at $K_s$\,=\,14.8\,mag, which includes all variable AGB sources identified in \citet{2015MNRAS.447.3973J} and thermally pulsing AGB models from \citet{2021ApJ...907...18R}.

\subsubsection{Red giants and supergiants}
\label{sssec:TrainEvolved}

RGB stars are a significant population that exhibit similar colours and magnitudes to faint YSOs in M\,33. We began with M-type stars identified in \citet{2005AJ....129..729R} as described in subsection Sect.\,\ref{sssec:TrainAGB}. Sources were rejected from the RGB sample if their $K_s$-band magnitude was brighter than the TRGB magnitude \citep[$K_s$\,=\,18.11\,mag,][]{2021ApJ...907...18R}. Since RGBs and Galactic foreground sources overlap in colour-space at fainter magnitudes \citep[e.g.][]{2021MNRAS.507.5106K}, we reject any source with a \textit{Gaia} proper motion consistent with a Galactic star (see Sect.\ref{sssec:TrainFG}). A colour cut was made at $J$\,$-$\,$K_s$\,>\,0.8\,mag to remove spurious near-IR matches; this value was selected based on the {\sc{TRILEGAL}} \citep{2005A&A...436..895G} Galactic foreground simulation mentioned in Sect.\ref{sssec:TrainFG}.

The \citet{2005AJ....129..729R} sample includes only RGBs brighter than $K$\,$\sim$\,18.9\,mag. Therefore the sample was augmented with additional spectroscopically confirmed fainter RGB sources from NGC\,6822 ($\mu{_{_{\rm NGC6822}}}$\,=\,23.34\,mag, \citealt{2019MNRAS.490..832J}; $\mu{_{_{\rm M33}}}$\,=\,24.67\,mag, \citealt{2014AJ....148...17D}), using the RGB training set compiled in \citet{2021MNRAS.507.5106K}. The process by which RGBs from both galaxies are combined to form the training class is discussed further in Section\,\ref{ssec:TrainDown}.

RSG stars are a young population \citep[$\sim$\,10\,$-$\,30\,Myrs,][]{2019A&A...624A.128B} which may contaminate the brighter end of a YSO sample. They can be dusty and due to their relative youth are located near sites of star formation \citep[e.g.][]{2020ApJ...892...91H,2021MNRAS.507.5106K}. RSGs were identified from optical and IR photometry using machine learning techniques in \citet{2022arXiv220308125M}, however as these sources lack further confirmation, such as spectroscopy, they are not included in the training set in order to maintain its purity. We adopt spectroscopically confirmed RSGs from the catalogue of \citet{2016AJ....152...62M}, confirmed based on their radial velocities and the presence of a strong Ca\,{\sc ii} triplet in their spectra. Using the RSG training set employed in NGC\,6822 \citep{2021MNRAS.507.5106K} as a guide, colour cuts at 0.4\,<\,$J$\,$-$\,$K_s$\,<\,2.5\,mag were made to remove a small number of spurious near-IR matches. 

\subsubsection{Blue stars}
\label{sssec:TrainBlue}

We include a class for bright and bluer stellar sources in M\,33. These include bright main-sequence stars as well as other classes not numerous enough to warrant a separate class; these are labelled collectively as `blue stars' (BS) in our classification scheme. These stars represent a younger population in M\,33 compared to e.g., AGB or RGB classes. A machine learning based, photometric identification of these populations is presented in \citet{2022arXiv220308125M} however; as with RSGs (see Sect.\,\ref{sssec:TrainEvolved}) we cannot utilise their catalogues to populate our BS class due to the lack of higher level classification.

The BS class is populated with spectroscopically confirmed O-, B- and A-type main-sequence stars from the catalogues of \citet{2016AJ....152...62M}. Main-sequence stars were sorted into their spectral types based on the relative strengths of Balmer lines (H$\delta$, H$\gamma$ and H$\beta$) and the presence and ratio of He lines. Additionally we include sources they classified as Luminous Blue Variables (LBV), yellow super giant stars (YSGs) and H\,{\sc ii} regions. \citet{2016AJ....152...62M} separate LBVs from unresolved H\,{\sc ii} regions based on the presence of strong Fe\,{\sc ii} lines \citep{2007AJ....134.2474M}. YSGs were identified using radial velocities and the presence of the O\,{\sc i} triplet at $\lambda$\,$\sim$\,777.4\,nm, to separate YSGs from foreground yellow dwarfs \citep{2012ApJ...750...97D}. Further colour cuts are set at $-$0.5\,<\,$J$\,$-$\,$K_s$\,<\,0.3 mag.

\subsubsection{Wolf-Rayet stars}
\label{sssec:TrainWR}

Wolf-Rayet (WR) stars are a relatively rare population with only $\sim$200 confirmed across the disk of M\,33 \citep{2011ApJ...733..123N}; they can present near-IR colours similar to those of YSOs and are often located close to regions of ongoing star formation \citep{2007AJ....134.2474M,2012AJ....143...43F}. Therefore, WR stars can contaminate YSO samples and are included in our classification scheme. The WR training set is comprised of spectroscopically confirmed sources from the catalogues of \citet{2016AJ....152...62M} and \citet{2011ApJ...733..123N}.

\subsubsection{Young stellar objects}
\label{sssec:TrainYSOs}

Our training sample of YSOs contains sources from both the Magellanic Clouds and NGC\,6822. YSOs with scaled near-IR magnitudes brighter than the detection thresholds in Sect.\,\ref{ssec:NIRData} were selected from catalogues of spectroscopically confirmed YSOs, \citet{2013MNRAS.428.3001O} for the SMC and \citet{2017MNRAS.470.3250J} for the LMC. Near-IR data for these sources were transformed from the native IRSF photometric system \citep{2007PASJ...59..615K} to the WFCAM photometric system as detailed in \citet{2021MNRAS.507.5106K}. This resulted in 69 LMC and 26 SMC sources for the YSO training class. We further include 55 YSOs in NGC\,6822. These YSOs were first identified in \citet{2019MNRAS.490..832J} and \citet{2020ApJ...892...91H} using mid-IR photometry and spectral energy distribution (SED) fitting with evolutionary models \citep{2006ApJS..167..256R, 2017A&A...600A..11R}, and confirmed using machine learning techniques \citep{2021MNRAS.507.5106K}. 

\subsection{Down-sampling of large training classes}
\label{ssec:TrainDown}

When one or more particularly numerous classes dominate the training set, the classifier training is faced with many more examples of those classes to the detriment of sparser classes. Hence the balance of class sizes in the training set affects classifier performance \citep[e.g.][]{4410397,8122151}. Due to real astrophysical population differences as well as the varied selection methods, the number of sources available for each class vary from 85 for WR to $\sim$\,7000 for FG. To ensure the PRF has the highest possible accuracy across all classes it was necessary to down-sample the four most numerous training set classes, FG, RGB, OAGB and CAGB. The positive effect of the down-sampling on classifier accuracy is shown in Sect.\,\ref{sssec:ConfMat}.

\begin{table}
\caption{Number of sources for each class for the five training sets (see Sect.\,\ref{ssec:Training Set}) after down-sampling of large training classes (see Sect.\,\ref{ssec:TrainDown}).} 
\centering
\begin{tabular}{lc}
\hline
PRF class & Number TS Sources \\ 
\hline 
YSO  & \llap{1}50    \\
OAGB  & \llap{1}72   \\
CAGB & 91    \\
AGN  & \llap{1}25  \\
FG  & \llap{2}83 \\
RGB  & \llap{2}00  \\
RSG  & \llap{1}80     \\
BS  & \llap{3}47   \\
WR & 85  \\
\hline
Total Sources& \llap{16}31 \\
\hline
\end{tabular}

\label{tab:TrainData}
\end{table}

The RGB training sources come from two sets of data, one in M\,33 and another from NGC\,6822 (see Sect.\,\ref{sssec:TrainEvolved}). Given the very different properties of these two galaxies (namely in terms of total stellar mass: $M_*{_{_{\rm NGC6822}}}$\,$\sim$\,1.5\,$\times$\,10$^8$\,M$_\odot$, \citealt{2014PASP..126.1079M}; $M_*{_{_{\rm M33}}}$\,$\sim$\,5.5\,$\times$\,10$^9$\,M$_\odot$, \citealt{2014A&A...572A..23C,2017AJ....154...41K}), and vastly different source density in CMD/CCD parameter space of confirmed RGB sources in each galaxy, these two RGB populations cannot just be added without introducing non-astrophysical biases that would affect the classifier performance. It was therefore necessary to down-sample the RGB sample from M\,33 to be more comparable with that of RGBs from NGC\,6822. This was done by comparing the fraction of NGC\,6822 RGBs above and below the M\,33 sample cut-off when scaled to the same distance (see Sect.\,\ref{sssec:TrainEvolved}). Reducing the M\,33 RGB sub-sample by a factor of 1 in 24 provides homogeneity in the combined NGC\,6822 and M\,33 RGB sub-samples across the M\,33 RGB cut off. For simplicity, the same down-sampling factor was applied to the other large classes (FG, OAGB and CAGB).

It is somewhat inevitable that down-sampling of the large classes will introduce some stochastic selection effects. Such effects, as well as those resulting from the train/test splitting of the sample, are counteracted by repeating both down-sampling and train/test splitting multiple times. For classes which cover a large range of magnitudes such as the FG class we checked that the down-sampling still adequately samples the parameter space (see Fig.\,\ref{fig:FG_DS_CMD}). 

In total we performed the down-sampling of the four larger classes randomly five times to create different training sets for the PRF. This number was selected based on achieving a stable number of YSOs recovered in common with each down-sampled training set (see Appendix\,A online only material). The number of sources in each training set class are given in Table\,\ref{tab:TrainData}. Each training set was used to train a PRF classifier which was run 20 times with different random seeds for the train/test split, totalling 100 runs.

\section{Results}
\label{sec:Results}

Using the training set defined in Sect.\,\ref{ssec:Training Set} the PRF classifier is applied 100 times to the 162,746 sources remaining in the catalogue.

\subsection{Confusion matrices}
\label{sssec:ConfMat}

\begin{figure}
    \centering
    \includegraphics[width=8cm]{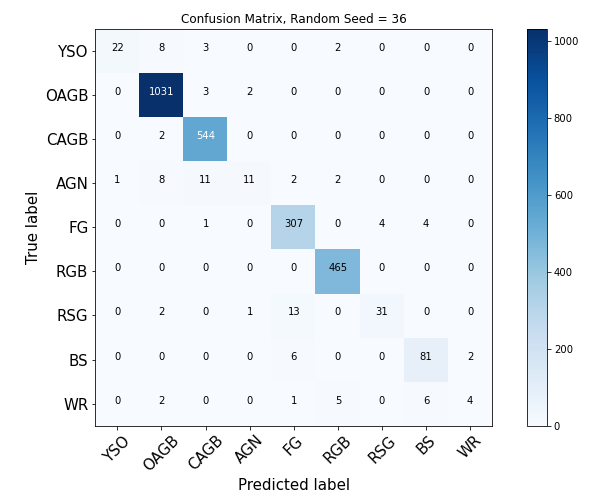}
    \includegraphics[width=8cm]{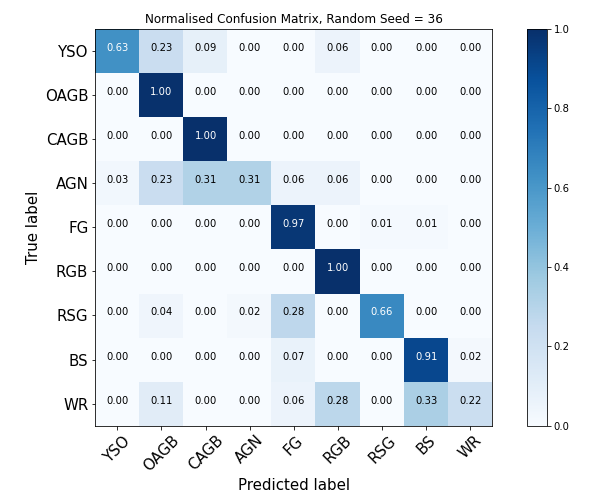}
    \caption{Non-normalised (top) and normalised (bottom) confusion matrices for an example PRF run with no class down-sampling (see text). The large classes achieve high accuracy, however for the smaller classes high levels of confusion are evident.}
    \label{fig:CM_noDS}
\end{figure}

Confusion matrices provide a helpful visualisation of the classifier's accuracy. Each matrix shows the PRF classification of the 25\% of sources in the test set classified using the remaining 75\% of training set sources. In Fig.\,\ref{fig:CM_noDS} the confusion matrices, both non-normalised and normalised, show the accuracy of a PRF classifier using the training set without any down-sampling applied. High classification accuracy is achieved for the large classes to the detriment of all other classes: sources from the smaller classes are often misclassified into the four large classes. In particular for YSOs, without down-sampling the PRF achieves accuracies ranging from 55 to 75\,per\,cent across the 100 runs with a median value of 66.5\,per\,cent. 

\begin{figure}
    \centering
    \includegraphics[width=8cm]{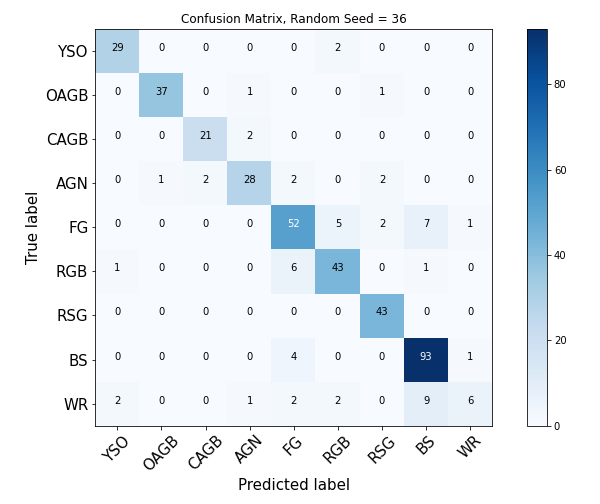}
    \includegraphics[width=8cm]{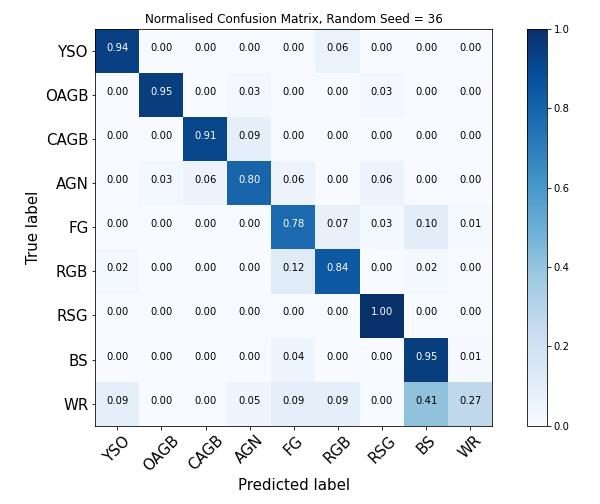}
    \caption{Non-normalised and normalised confusion matrices (respectively top and bottom) for the PRF run using the same random seed as those shown in Fig.\ref{fig:CM_noDS}, but here with class down-sampling (see text). The misclassifications for the smaller classes are very effectively reduced. }
    \label{fig:ConfDS}
\end{figure}

In Fig.\,\ref{fig:ConfDS} we show the PRF matrices, using the same random seed as those shown in Fig.\,\ref{fig:CM_noDS}, with down-sampling applied as described in Sect.\ref{ssec:TrainDown}. In general an improvement in the overall PRF classification accuracy, exemplified by the strong diagonal feature in the normalised matrix, is evident. In particular the YSO classification accuracy significantly improves, ranging from 62 to 97\,per\,cent, with a median value of 82\,per\,cent across all runs. Across the 100 PRF runs the median class-averaged accuracy is 87\,per\,cent. The estimated accuracy per PRF is skewed by the WR class which performs significantly worse than all others by a large margin (see Fig.\,\ref{fig:ConfDS}); we discuss source misclassification and contamination in the following section.

\subsubsection{Potential misclassifications and class contamination}
\label{sssec:misclass}

As already mentioned, YSOs in the training set are recovered with high accuracy (median accuracy of 82\,per\,cent). More specifically 67 PRF runs achieve an YSO accuracy of over 80\,per\,cent, and only 4 runs have accuracy below 70\,per\,cent. Misclassified training set YSOs are most often placed into the OAGB, RGB and WR classes. Some OAGB, RGB and dusty WR stars have similar near-IR colours and magnitude to YSOs which is the likely cause for the confusion in the PRF's classification. Additionally WR stars are likely to be associated with sites of bright far-IR emission \citep[e.g.][]{2012AJ....143...43F} similar to YSOs. 

The YSO class suffers from very low levels of contamination from other classes; the highest fraction of incorrectly classified YSOs in the test sample are WRs due to the similarities noted above. Dusty WRs are however relatively rare therefore the absolute contamination of YSOs remains very low. The opposite happens for the RGB class: their fractional contamination to the YSO class is low however they are very numerous, meaning RGBs can still be important contaminants of the YSO sample. We use training set sources that after down-sampling are returned to the main catalogue to further investigate YSO contamination in the final classifier output (Sect.\,\ref{sssec:ClassOut}).

As noted previously, WR is the worse performing class. This class has the fewest training sources available (85 sources), and is misclassified into AGN, BS, FG, RGB and YSO classes. Of these the BS class is the dominant misclassification, in some runs even out-scoring the correct classification (see Fig.\,\ref{fig:ConfDS}). The lower performance of the PRF in WR classification is a consequence of previously discussed similarities to other classes and the small training set size for this class.

For the AGNs we see some confusion with the OAGB and CAGB classes, likely due to the fact that AGN can have near-IR colours similar to those of the AGB populations \citep{2011A&A...531A.137H,2022MNRAS.tmp.2008P}. A similar effect was also seen in AGN classifications behind NGC\,6822 \citep{2021MNRAS.507.5106K}.

\subsection{Final classifier outputs}
\label{sssec:ClassOut}

Each of the individual 100 PRF runs provides a classification for all sources not included in the training/testing sets. These 100 classifications provide a score between 0 and 100 for each source and for each class, $n_{\rm class}$ with class\,=\, YSO, FG, etc. The PRF classifies 41\,per\,cent of sources into the same class over all runs (i.e. $max(n_{\rm class})$\,=\,100). These sources are included in our subsequent analysis, and are henceforth referred to as classified. The breakdown of the 66,378 classified sources into the different PRF classes is given in Table\,\ref{tab:Classes}. 

\begin{table}
\caption{Number of sources in M\,33 classified into each PRF class and total number sources including those from the training set after down-sampling of the largest classes (see Sect.\,\ref{ssec:TrainDown}).} 
\centering
\begin{tabular}{lcc}
\hline
PRF Class &  Classified & Training \& classified \\ 
                   &  sources & sources \\
\hline 
YSO  & \llap{49}85 &  \\
OAGB  & \llap{182}14 &  \llap{18}387\\
CAGB & \llap{20}86 & \llap{2}177\\
AGN  & \llap{37}57 & \llap{3}793\\
FG  & \llap{52}94 & \llap{5}577\\
RGB  & \llap{274}22 & \llap{274}98\\
RSG  & \llap{14}24 & \llap{1}604\\
BS  & \llap{31}11 & \llap{3}458\\
WR & 82 & 167\\
\hline
\end{tabular}
\label{tab:Classes}
\end{table}

\begin{figure*}
    \centering
    \includegraphics[height=22.5cm]{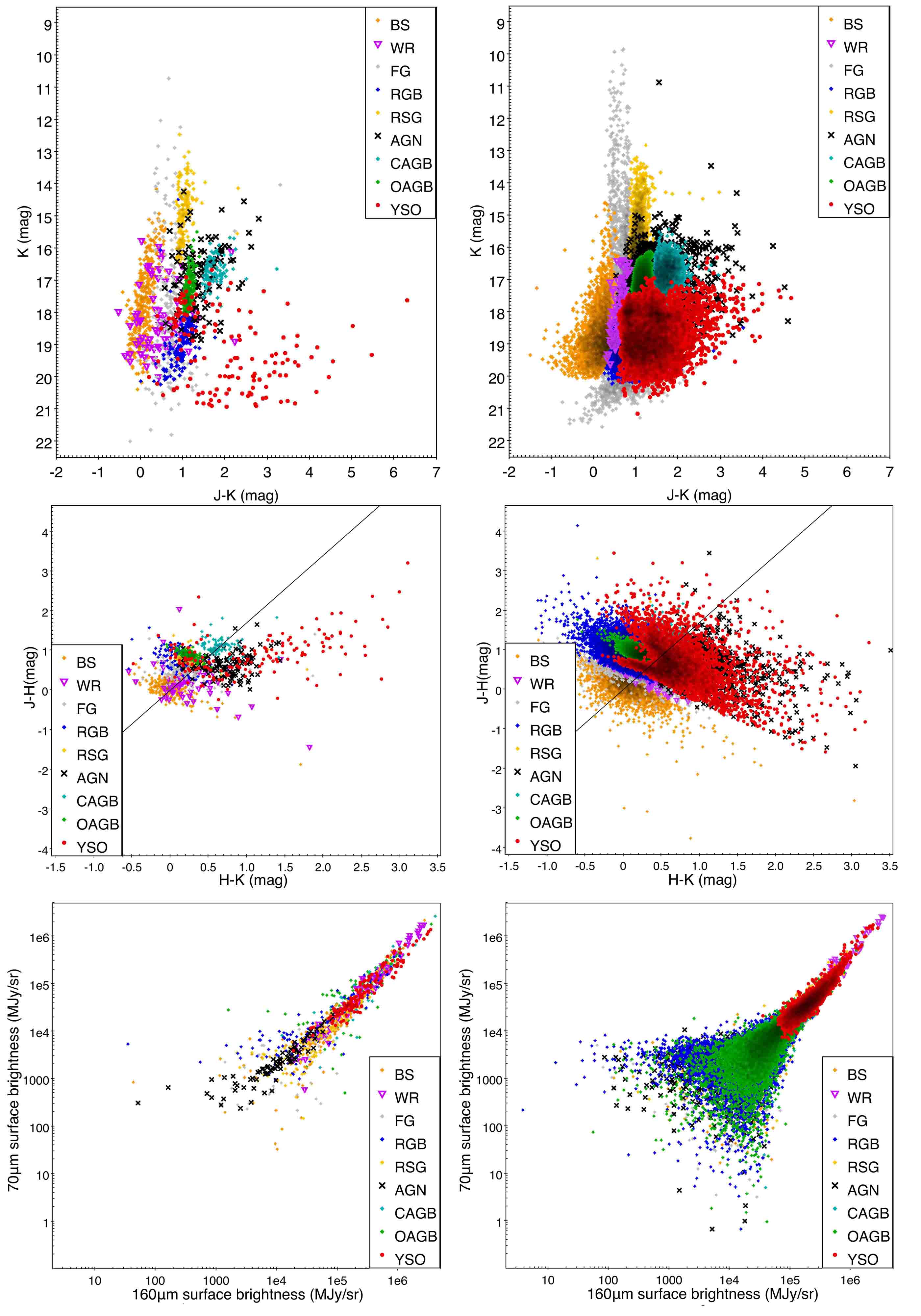}
    \caption{CMD, CCD and far-IR brightness plots of the training set sources (left) and for the classified sources (right). Colour-coding is given in the legend. The reddening line shown in the CCD plots is derived using the coefficients from \citet{1985ApJ...288..618R}.} 
    \label{fig:CompPlots}
\end{figure*}

In Fig.\,\ref{fig:CompPlots} we present a CMD, colour-colour diagram (CCD) and far-IR brightness plot for both training/testing set data and classified sources. The plots show that, for every class, training and classified sources occupy a similar position in parameter space. Whilst both training and classified YSOs cover a similar range of $J$\,$-$\,$K_s$ colours from 0.5 to 5\,mag, and $K_s$-band magnitudes from 16 to 21\,mag, at magnitudes fainter than $K_s$\,$=$\,19.5\,mag classified YSOs are seldom redder than $J$\,$-$\,$K_s$\,$=$\,2.5\,mag. This is primarily due to fact that some training set YSOs can have $J$- and $H$-band magnitudes fainter than the near-IR catalogue's detection thresholds (see Sect.\,\ref{ssec:NIRData}). This arises from practical considerations in the design of the near-IR observations, with shorter wavelength images not deep enough to characterise the redder sources, being these YSOs or AGBs. Therefore the faintest YSOs we identify are not particularly red, and, as expected, no classified YSOs are found outside the colour and magnitude ranges described by the training set YSOs.

Figure\,\ref{fig:CompPlots} also shows that whereas in the training set there is a region of the CMD occupied by both OAGBs and CAGBs brighter than $K_s$\,=\,16\,mag, in the classified sources this region is dominated by AGN classifications. These sources are likely missclassified due to the confusion between these classes commented upon in Sect.\,\ref{sssec:misclass}.

We discussed potential YSO contamination in Sect.\,\ref{sssec:misclass}. The confidence matrices however only provide the likelihood of contamination for a single PRF run; for a source to effectively become a contaminant of the YSO class, it needs to be consistently classified in that class 100 times. We use the sources from the training set that are returned to the catalogue for classification to quantify such effects for the most numerous astrophysical classes. In total 655 sources are excluded from the RGB, FG, OAGB and CAGB training sets after down-sampling (see Sect.\,\ref{ssec:TrainDown}). These sources with known classification are used to provide an additional estimate of class contamination alongside the statistics provided by the confusion matrices (Sect.\,\ref{sssec:ConfMat}). Of these 655 sources, 358 ($\sim$\,55\,per\,cent) are classified by the PRF, with 330 assigned to the correct literature class (i.e. 92\,per\,cent of classified sources are correctly classified). The 28 incorrectly classified sources (seventeen OAGBs, three CAGBs and twelve FGs) are misclassified as sixteen RSGs, eight AGNs and four BSs. None of these sources are classified as a YSO. Noteworthy is the fact that despite the considerations discussed in Sect.\,\ref{sssec:misclass}, none of the RGBs are misclassified, as YSO or any other class.

\subsection{Spatial distributions}
\label{sssec:SpDist}

As noted in Sect.\,\ref{ssec:NIRData} sensitivity issues become apparent for $K_s$\,$>$\,19.2\,mag. The effects of source crowing increase significantly towards the centre of the galaxy (central $\sim$\,7\,$\times$\,7\,arcmin$^2$ region), since evolved star density profiles decrease as a function of radial distance \citep[e.g.][]{2005AJ....129..729R,2021ApJS..253...53W}. Due to crowding the PRF's classifications are less certain in the central region, with a larger fraction of sources being assigned $n_{\rm class}$\,<\,100, effectively remaining unclassified by the PRF. In the central region 30\,per\,cent of sources are classified compared to 42\,per\,cent in the outer regions. While crowding affects the identification for all classes, classes dominated by fainter sources are more severely affected.

\begin{figure*}
    \centering
    \includegraphics[width=5.91cm]{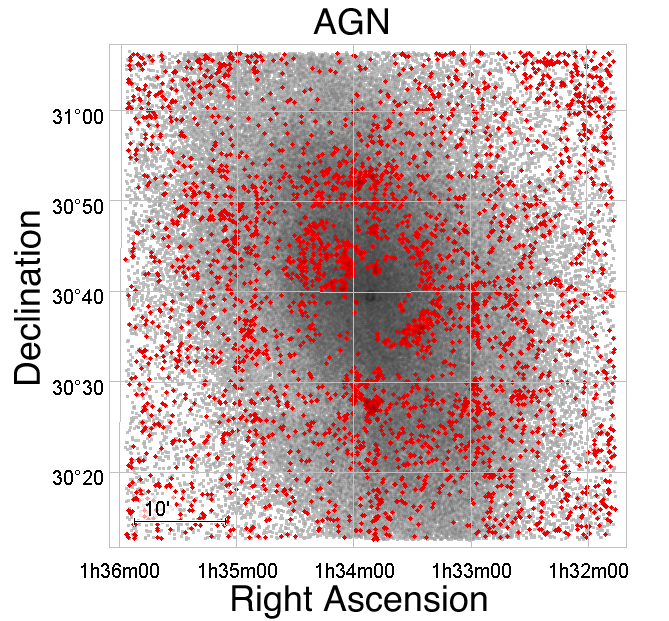} 
    \includegraphics[width=5.91cm]{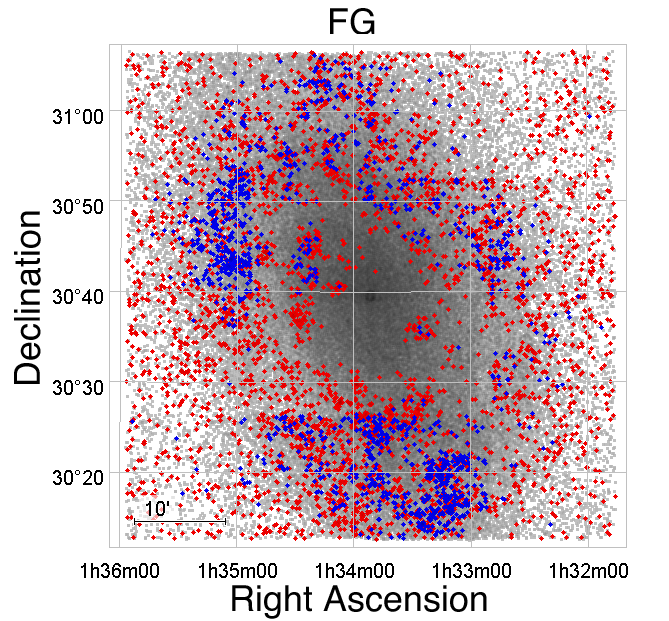}
    \includegraphics[width=5.91cm]{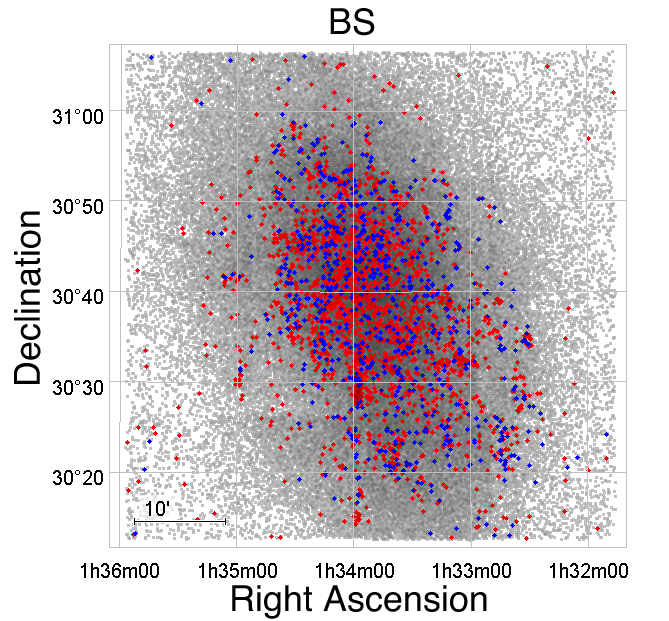}
    
    \includegraphics[width=5.91cm]{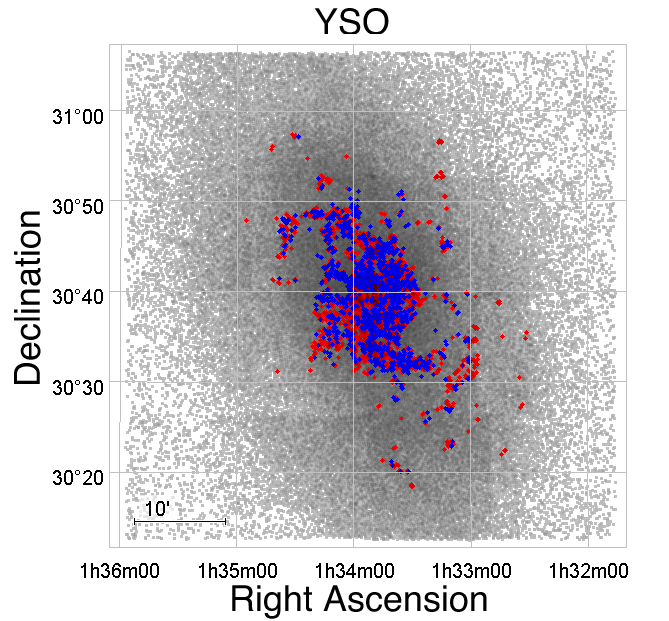}
    \includegraphics[width=5.91cm]{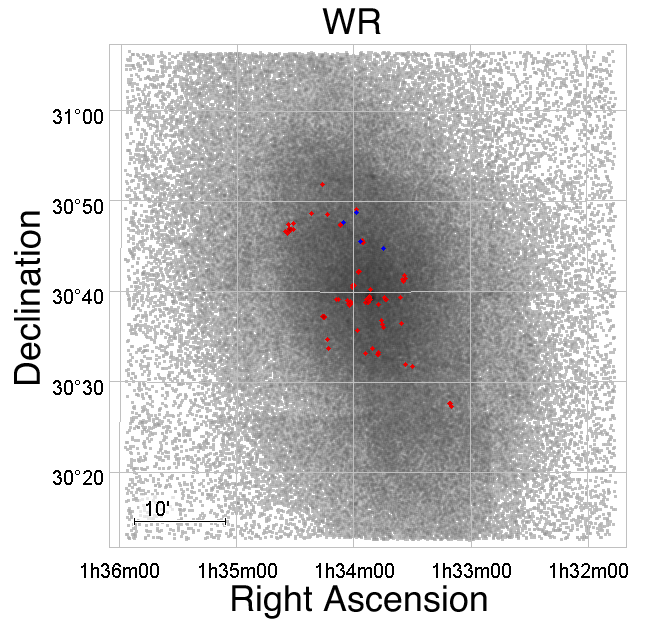}
    \includegraphics[width=5.91cm]{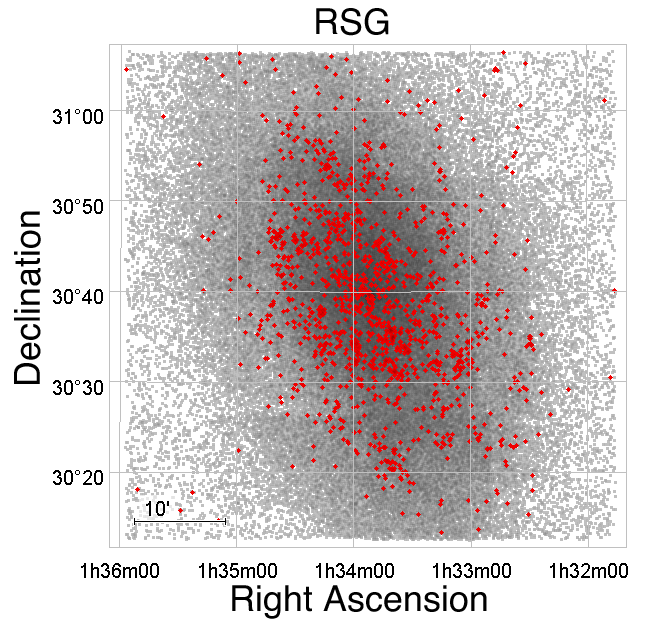}

    \includegraphics[width=5.91cm]{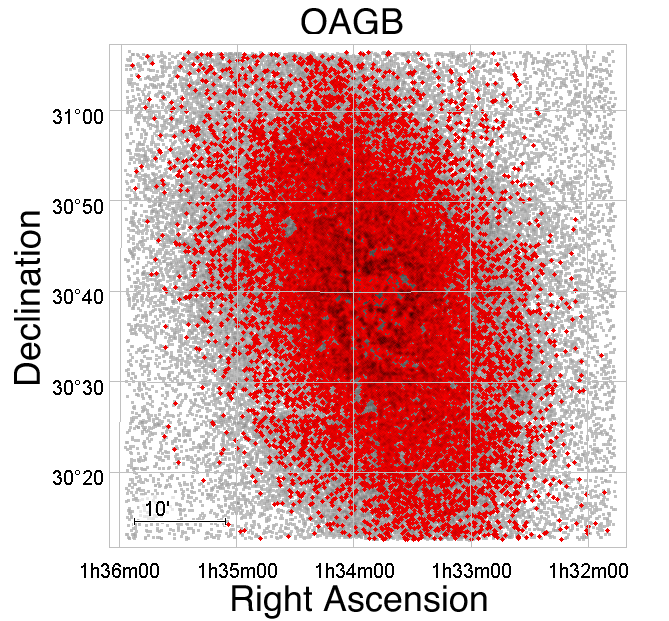}
    \includegraphics[width=5.91cm]{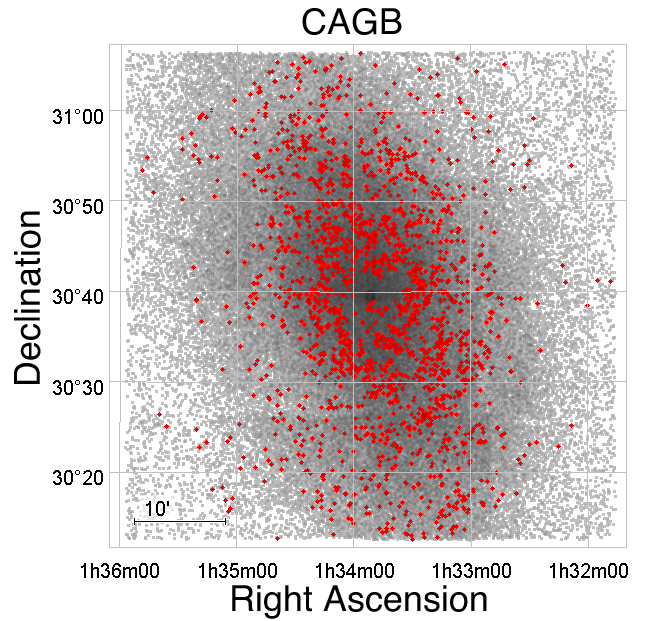}
    \includegraphics[width=5.91cm]{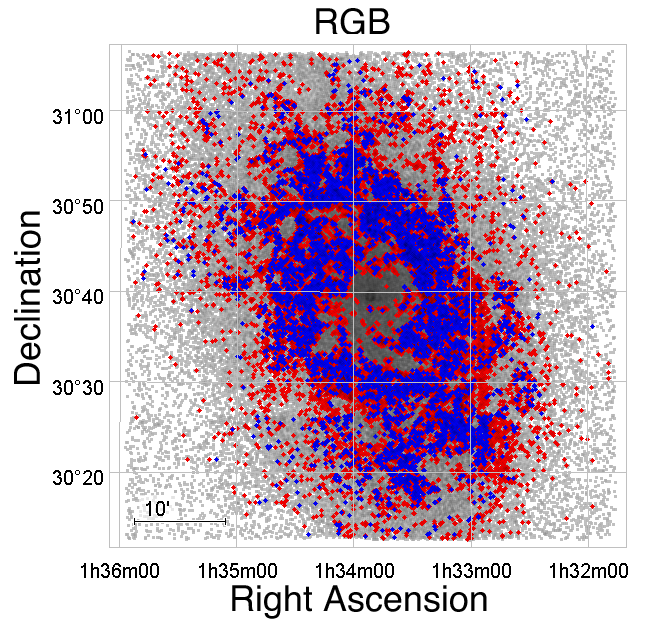}
    \caption{Spatial distributions for each PRF class. Sources with $K_s$\,$<$\,19.2\,mag  and $K_s$\,$\geqslant$\,19.2\,mag are shown respectively in red and blue. The full catalogue is shown in the background.}
    \label{fig:SpatDist}
\end{figure*}

In Fig.\,\ref{fig:SpatDist} we show the spatial distributions of classified sources for each class. We briefly highlight some salient features of non-YSO distributions, however a thorough discussion is beyond the scope of this paper. We discuss the YSO distribution in Sect.\,\ref{ssec:M33Structure}. 

AGN and FG sources are fairly evenly distributed across the field as expected. AGNs are not identified in the crowded central region of M\,33, since the increased point-source density and brighter completeness limit there make it very difficult to identify background sources. Furthermore, as noted in Sects.\,\ref{sssec:misclass} and \ref{sssec:ClassOut} there is some confusion between the AGN and AGB classes. These effects are most apparent in the centre of M\,33 where the AGN distribution appears less uniform than in the outer regions. The FG class shows some correlation with the overall catalogue source density outside the centre of M\,33 especially at fainter magnitudes. This behaviour is reversed in the central region where FG sources are seldom classified, consequence of the crowding and associated completeness issue. 

The AGB and RGB classes show distributions throughout the disk of M\,33 in agreement with the source density distributions previously reported \citep{2015MNRAS.447.3973J,2021ApJS..253...53W}. We recover the faint two arm morphology seen in the RGB and combined OAGB and CAGB distributions in the inner $\sim$\,20\,$\times$\,10\,arcmin$^2$ region \citep{2021ApJS..253...53W}. The CAGB class does not exhibit a source density increase towards to the centre of M\,33, as is seen in the OAGBs, in agreement with the density profiles observed by \citet{2005AJ....129..729R}. The strong ring-like CAGB structures (at $\sim$\,3.5\,kpc from the centre of M\,33, \citealt{2004A&A...425L..37B,2007A&A...471..467B}) are not seen in our analysis. As already mentioned, in the central region, crowding affects the PRF classification, with fewer classified faint sources present, as seen in particular for the RGB distribution.

The BS and RSG classes represent stellar populations younger than AGB and RGB classes. Their distributions are highly structured, more closely associated with the spiral arms. For the BS class this morphology is in general agreement with the distribution of the young main-sequence population in the central region of M\,33 \citep[][MS distribution in their figure 22]{2021ApJS..253...53W}. The RSG distribution closely resembles that found by \citet[][see their figure 11]{2021AJ....161...79M} and \citet[][see their figure 11]{2021ApJ...907...18R}. The WR source distribution, even though very sparse, loosely follows the distribution of YSOs (Sect.\,\ref{ssec:M33Structure}).

\subsection{YSO distribution and clustering}
\label{ssec:YSOclusters}

\begin{figure}
    \centering
    \includegraphics[width=\columnwidth]{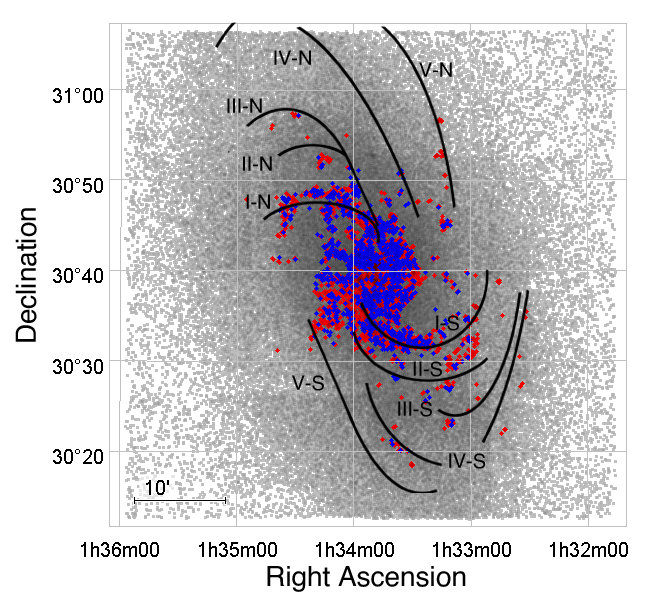}
    \caption{YSO distribution in M\,33, with the spiral structure adapted from \citet{1980ApJS...44..319H} overlaid (colour-coding as in Fig.\,\ref{fig:SpatDist}).}
    \label{fig:YSOspiral}
\end{figure}

The PRF identifies 4985 YSOs across the disk of M\,33; their properties are listed in Table\,\ref{tab:YSOs} and their distribution is shown in Fig.\,\ref{fig:YSOspiral}. As already discussed, the PRF classifies $\sim$\,30 to $\sim$\,42 per cent of sources in the catalogue; therefore this YSO sample is robust but unlikely to be complete. The YSO sources are found mostly in the central region of the galaxy and on the two major spiral arms of M\,33 (I-N and I-S). Arms I-N and I-S contain $\sim$\,300 YSOs each, with a similar total YSO mass (see Sect.\,\ref{ssec:YSOprop} for details on YSO mass estimates). The area adjacent to the base of I-S in which many YSOs are found is the base of arm IV-S \citep{1980ApJS...44..319H}. A small number of YSOs lie further along the other spiral arms. 

We identify SFRs in M\,33 by examining the spatial clustering of classified YSOs. These YSO clusters were identified using a Density-Based Spatial Clustering of Applications with Noise \citep[{\sc DBSCAN},][]{10.5555/3001460.3001507}. {\sc{DBSCAN}} is a clustering algorithm which finds density-based associations in spatial data. This process was performed using deprojected coordinates (see Appendix\,A, online only material, for details).

{\sc{DBSCAN}} requires two parameters that can be tuned to the data: a minimum number of YSOs in a cluster and a distance parameter $\epsilon$, the furthest distance at which a neighbour is selected. The minimum YSO number is set to eight, selected to avoid splitting the most apparent clusters and consistent with the value used in a similar analysis in NGC\,6822 \citep{2019MNRAS.490..832J}. We optimised the choice of $\epsilon$ using a k-nearest neighbours (k-NN) method. It analyses the distances between individual YSOs and finds the ``elbow-point'' in the distance distribution which is the optimal value for $\epsilon$ \citep{Rahmah_2016}. 

The initial run of {\sc{DBSCAN}} ($\epsilon$\,=\,0.1551) identified 23 clusters but was unable to identify clusters in the central region of M\,33 where the source density is much higher. To recover additional clusters, the process was repeated with progressively smaller $\epsilon$ values using those YSO sources that remained unassigned (see Table.\,\ref{tab:DBepsilon}). This process was repeated five times, after which the $\epsilon$ distance returned by the k-NN analysis effectively plateaued. Overall, {\sc DBSCAN} identifies 62 YSO clusters.

\begin{figure}
    \centering
    \includegraphics[width=\columnwidth]{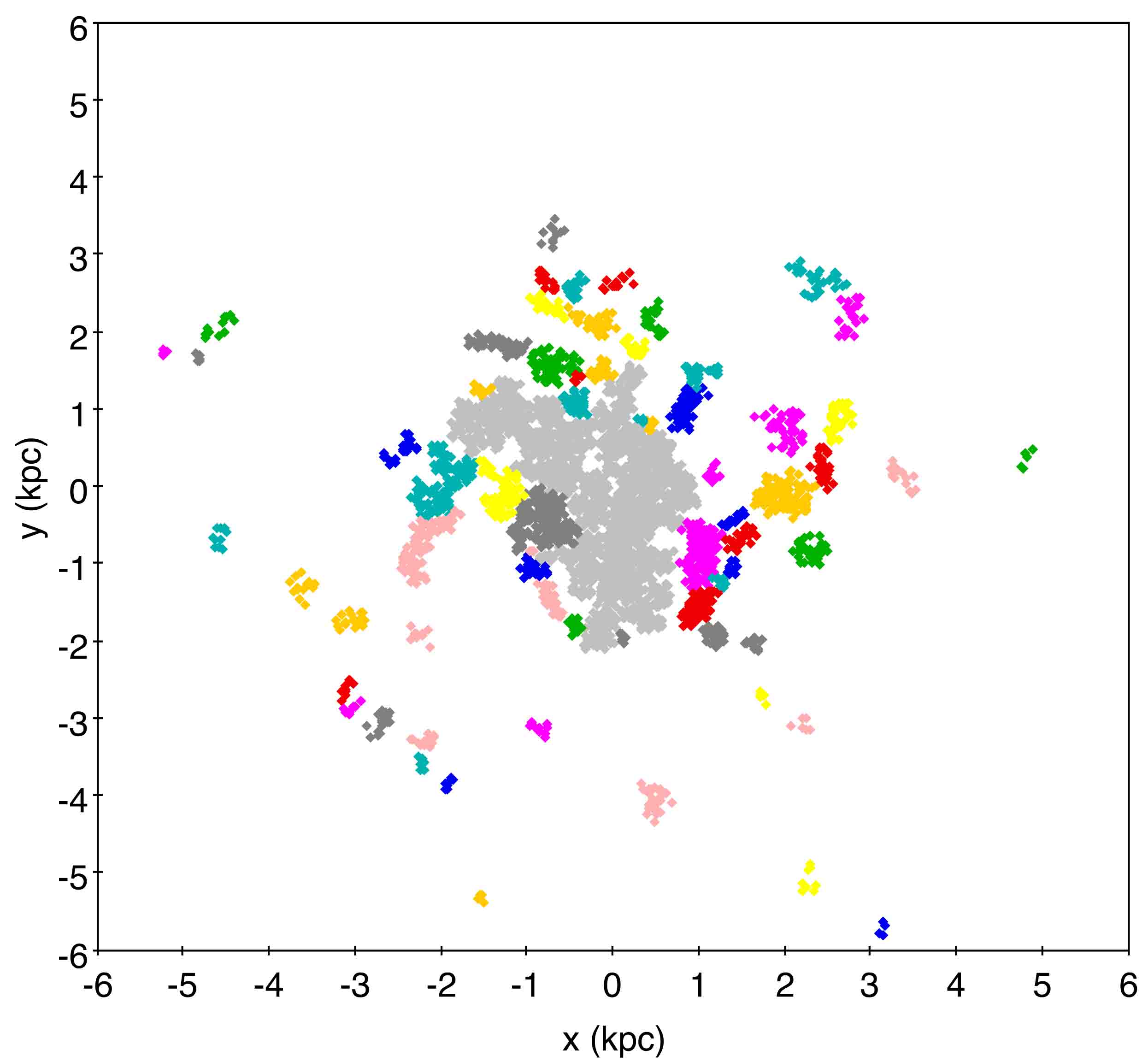}
    \caption{Clusters of YSOs identified by {\sc{DBSCAN}}, displayed in deprojected coordinates. The central region (see text) without identified clusters is shown in light grey colour. This projection is rotated by 90\,degree clockwise with respect to the sky coordinates shown in Fig.\,\ref{fig:YSOspiral}.}
    \label{fig:clusterXY}
\end{figure}

A visual inspection of the YSO source distribution revealed a small number of additional YSO clusters that did not meet the {\sc{DBSCAN}} criteria. One example is the H\,{\sc ii} region IC\,133, which has many indicators of massive star formation such as H$_2$O and OH maser emission \citep[respectively][]{1977A&A....54..969C,1987MNRAS.226..689S}, but was not identified by {\sc{DBSCAN}} due to its nine YSOs being spread across a larger area (131\,pc or 32\,arcsec). Six more clusters were identified by eye. A total of 68 YSO clusters (henceforth referred to as SFRs) were identified across the disk of M\,33, ranging in size from 31 to 550\,pc (7.5 to 132\,arcsec) and containing between 3 and 211 YSOs. The radii of the SFRs are broadly consistent, albeit at the higher end, with the GMC sizes in M\,33 analysed by \citet{2017A&A...601A.146C}; we discuss the relationship between SFRs and GMCs in Sect.\,\ref{sssec:GMC}. The SFR spatial distribution in deprojected coordinates is shown in Fig.\,\ref{fig:clusterXY}. The centre of each SFR is defined as the average of the members' positions and its radius is the largest distance from this average position. This definition of SFR size is consistent with that used by \citet{2017MNRAS.470.3250J} in NGC\,6822, allowing for a direct comparison of SFR properties in both galaxies (see Sect.\,\ref{ssec:M33Structure}). SFR properties are listed in Table\,\ref{tab:SFR}.

As discussed previously, in the central dense region of M\,33 {\sc DBSCAN} was unable to recover YSO clusters. In total 1986 YSOs were assigned to a SFR listed in Table\,\ref{tab:SFR}, 562 were unclustered and 2437 were left in the central dense ``remnant'' ($\sim$\,11.6\,$\times$\,10.4\,arcmin$^2$ or 2.8\,$\times$\,2.5\,kpc$^2$ in size, light grey in Fig.\,\ref{fig:clusterXY}). In general, the PRF works less well in this central region, with only 30\,per\,cent of sources classified as opposed to 41\,per\,cent overall. As already discussed, we identify fewer than expected RGB sources in this region (see Fig.\,\ref{fig:SpatDist} and Sect.\,\ref{sssec:SpDist}). Given their expected distribution in the M\,33 disk and strong overlap in colour-magnitude space with YSOs (see Fig.\,\ref{fig:CompPlots}), RGBs are an important contaminant class (see Sect\,\ref{sssec:misclass}), even if no known RGBs are misclassified as YSOs by the PRF (Sect.\,\ref{sssec:ClassOut}). Nevertheless, assuming in extremis that all 811 YSOs overlapping the RGB region of the CMD space are contaminants, we estimate that at most 30\,per\,cent of YSOs could be wrongly classified in the central region. We take this into account in the analysis in Sect.\,\ref{ssec:YSOprop}.

\begin{figure}
    \centering
    \includegraphics[width=\columnwidth]{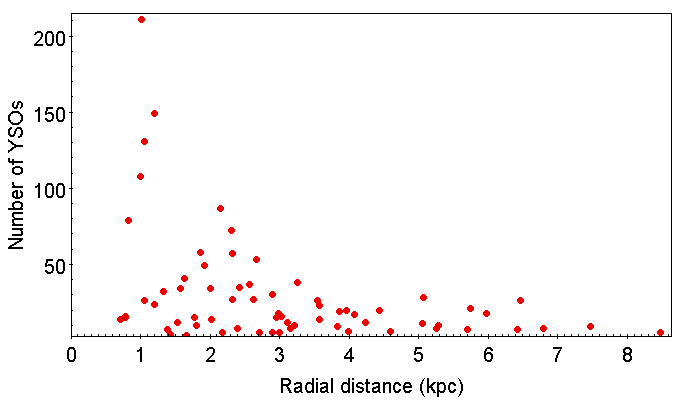}
    \includegraphics[width=\columnwidth]{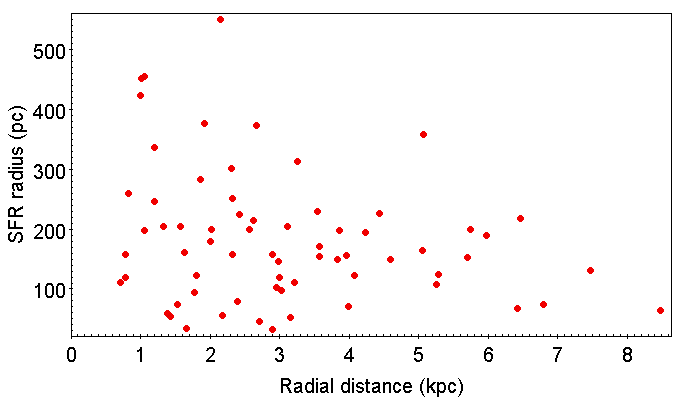}
    \caption{Number of YSOs (top) and radius (bottom) for each SFR identified by {\sc DBSCAN} as a function of radial distance. A decreasing profile with increasing distance from the centre is seen in both panels.}
    \label{fig:RadialPlots}
\end{figure}

In Fig.\,\ref{fig:RadialPlots} the number of YSOs per SFR and the size of each SFR are shown against the deprojected radial distance to the centre of M\,33: the largest and more numerous clusters are found closer to the centre of the disk.

\begin{table*}
    \caption{Catalogue of YSOs in M\,33 classified using the PRF analysis. For YSOs assigned to a SFR by the {\sc DBSCAN} analysis, the SFR ID is given. YSO mass estimates are discussed in Sect.\,\ref{ssec:YSOprop}. A sample of the table is provided here, the full catalogue is available as supplementary material.}
    \label{tab:YSOs}
    \centering
    \begin{tabular}{ccccccccccc}
    \hline
   RA\,(J2000) & Dec\,(J2000) & $J$ &  $J_{\rm err}$& $H$ & $H_{\rm err}$ & $K$& $K_{\rm err}$ & SFR &  mass \\
    h:m:s & deg:m:s & mag &  mag & mag & mag & mag & mag & ID & M{$_\odot$} \\
    \hline 
         01:33:49.16 & +30:40:17.7 & 18.48 & 0.066 & 17.89 & 0.047 & 16.99 & 0.051 &    & 13.9 \\
         01:34:10.32 & +30:36:40.7 & 19.17 & 0.061 & 18.28 & 0.078 & 17.28 & 0.057 & 26 & 12.9 \\
         01:34:06.18 & +30:37:47.3 & 19.03 & 0.054 & 17.77 & 0.050 & 16.35 & 0.039 & 39 & 19.8 \\
         01:33:48.66 & +30:44:48.3 & 19.48 & 0.143 & 18.55 & 0.107 & 18.04 & 0.087 & 48 & 20.1 \\
         01:33:37.54 & +30:36:02.1 & 21.13 & 0.260 & 20.25 & 0.306 & 19.85 & 0.318 & 56 & \,\,9.4\\
    \hline
    \end{tabular}
\end{table*}

\begin{table}
\caption{$\epsilon$ distances used in the {\sc{DBSCAN}} clustering analysis and the cumulative number of clusters recovered after each step (see text).}
\centering
\begin{tabular}{cc}
\hline
$\epsilon$ &Identified \\
(kpc)&clusters\\
\hline
0.1551 & \llap{2}3  \\
0.1064 & \llap{4}1 \\
0.0885 & \llap{5}0  \\
0.0852 & \llap{5}8  \\
0.0824 & \llap{6}2  \\
\hline
\end{tabular}

\label{tab:DBepsilon}
\end{table}

\begin{table*}
    \caption{Catalogue of SFRs in M\,33 identified using {\sc DBSCAN}. The evolution score is discussed in Sect.\,\ref{sssec:SFRevol}. A sample of the table is provided here, the full version is available as supplementary material.}
    \label{tab:SFR}
    \centering
    \begin{tabular}{cccccccc}
    \hline
   SFR & RA\,(J2000) & Dec\,(J2000) & Maximum radius  &Median radius& YSO & Evolution & SFR\\
   ID & h:m:s & deg:m:s & pc & pc  & number & score & identifiers \\
    \hline 
          1  & 01:34:17.91 & +30:37:21.5 & 195 & 88 & 12 & \llap{$-$}0.239& \\
          2  & 01:33:10.89 & +30:29:56.6 & 198 & 92 & 19 & 0.746&\\
          3  & 01:34:35.62 & +30:45:59.3 & 357 & 193 & 28& 0.239&NGC604-S \\
          4  & 01:34:13.24 & +30:45:59.3 & 376 & 156 & 49 & 0.388&\\
          5  & 01:33:13.07 & +30:45:12.2 & 218 & 99 & 26& 0.209&\\
    \hline
    \end{tabular}
\end{table*}

\section{Discussion}
\label{sec:Disc}

\subsection{The star forming regions in M33}
\label{ssec:M33Structure}

In this section we discuss the observed properties of SFRs in M\,33 and discuss their evolutionary status, using SFRs in NGC\,6822 (analysed using similar methods) as a benchmark.

\subsubsection{SFR observed properties}
\label{sssec:SFObProp}

Integrated optical to far-IR brightnesses can be used to characterise and probe the activity in SFRs. H$\alpha$ emission in SFRs arises from unobscured massive YSOs and young massive stars, whilst emission at 24\,$\mu$m traces warm dust associated with recent star formation activity \citep[e.g.][]{2012ARA&A..50..531K}. In order for H$\alpha$ emission arising from massive young stars to be observed, sufficient time for the ionising radiation and winds of those stars to clear the surrounding, obscuring dust must have passed. Hence the ratio of H$\alpha$ to 24$\mu$m provides a measure of the levels of exposed to embedded star formation respectively \citep[e.g.][]{2017ApJ...835..278S} and from this the relative ages of SFRs can be estimated \citep{2019MNRAS.490..832J}. Recently, both H$\alpha$ and 24-$\mu$m emission have been used as indicators of youth in age estimations of stellar clusters (with ages $>$\,2\,Myr) across the disk of M\,33 \citep{2022AJ....163...16M}.

The ratio of far-IR emission observed with {\it Herschel} has been shown to spatially correlate with other shorter wavelength tracers of star formation across many nearby galaxies \citep{2010A&A...518L..61B} including in M\,33 \citep{2007A&A...466..509T,2010A&A...518L..67K}. Specifically, the ratio of 250-$\mu$m to 500-$\mu$m emission in H\,{\sc ii} regions across NGC\,6822 correlates well with other tracers of ongoing star formation \citep{2010A&A...518L..55G}, pinpointing SFRs analysed in detail in more recent studies \citep{2019MNRAS.490..832J,2021MNRAS.507.5106K}. Longer wavelength emission is especially valuable at tracing the earliest stages of star formation, in which light emitted at shorter wavelengths is either obscured by dust (e.g., H$\alpha$) or the dust has not been sufficiently heated to become bright at mid-IR wavelengths. 

Thus optical to far-IR emission can be expected to peak at different stages of the evolution of a SFR. A higher flux at longer wavelengths compared to H$\alpha$ suggests rising star formation activity \citep[e.g.][]{2019MNRAS.490..832J}; the opposite behaviour is expected for regions in which star formation is ending and exposed massive stars begin to move onto the main sequence \citep[e.g.][]{2003ARA&A..41...57L,2010ARA&A..48..431P}. Hence by comparing the ratios of H$\alpha$ to 24$\mu$m and 250$\mu$m to 500$\mu$m ([H$\alpha$]$/$[24$\mu$m] and [250$\mu$m]$/$[500$\mu$m] respectively) for several SFRs it is possible to establish their evolutionary sequence.

For each SFR identified by the {\sc{DBSCAN}} analysis background subtracted aperture photometry was performed in H$\alpha$, 24-$\mu$m {\it Spitzer}-{\sc MIPS}, 250- and 500-$\mu$m {\it Herschel}-SPIRE images (see Sect.\,\ref{ssec:AncData} for image details) to measure an average brightness within each aperture. The position and size of the apertures were set to the SFR centre and radius (see Table\,\ref{tab:SFR}). In order to calibrate the properties and evolutionary status of SFRs in M\,33 we used regions in NGC\,6822 that have been well-characterised in the literature \citep{2017ApJ...835..278S,2019MNRAS.490..832J,2021MNRAS.507.5106K} as a benchmark for which we performed similar measurements. Positions and radii for NGC\,6822 SFRs were taken from table 9 of \citet{2019MNRAS.490..832J}. These seven regions are the complete census of significant sites of star formation in NGC\,6822. We do not include in this analysis the smaller SFRs newly identified in \citet{2021MNRAS.507.5106K} since an established evolutionary sequence is not available for these regions.

\begin{figure}
    \centering
    \includegraphics[width=7.3cm]{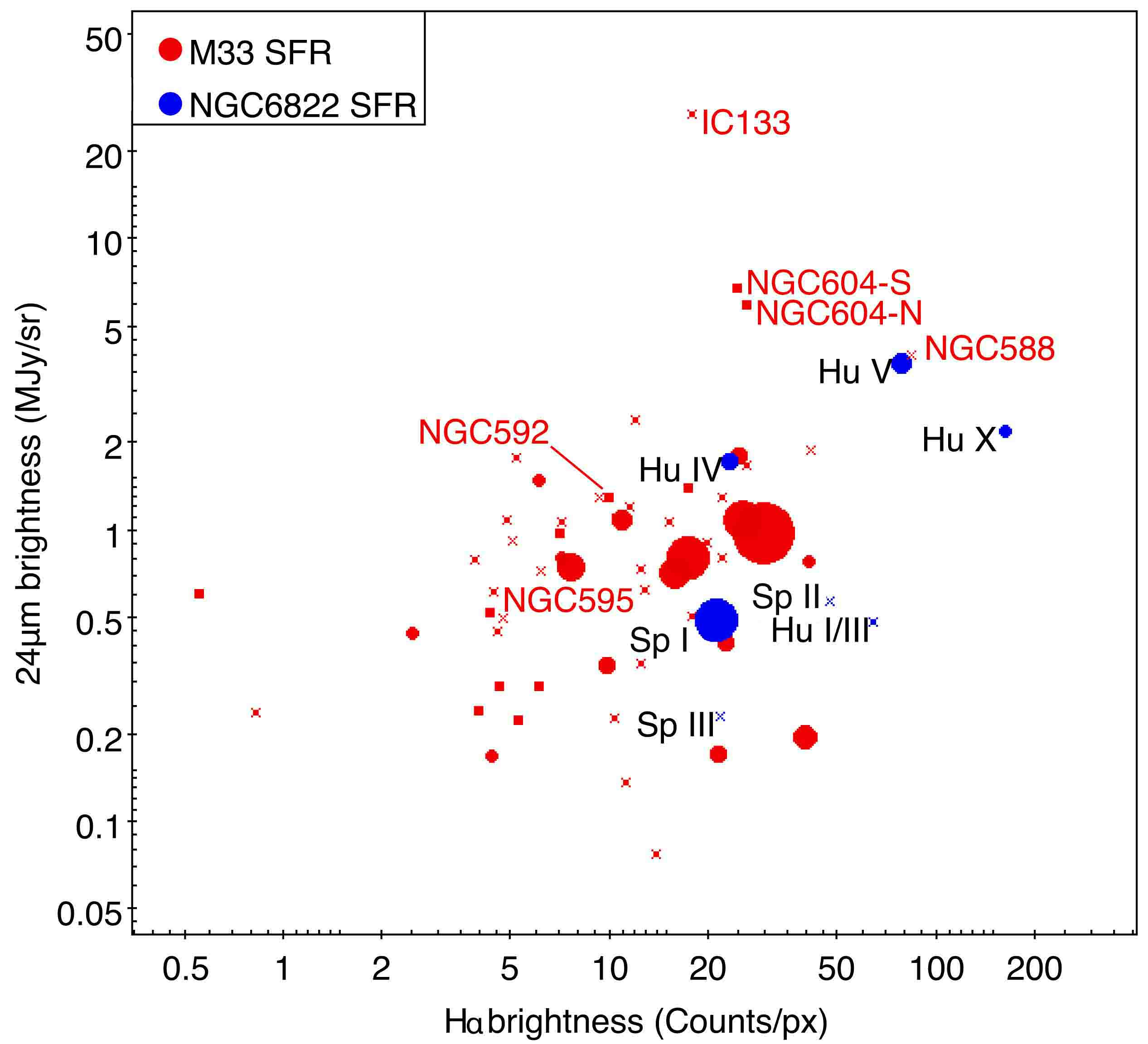}
    \includegraphics[width=7.4cm]{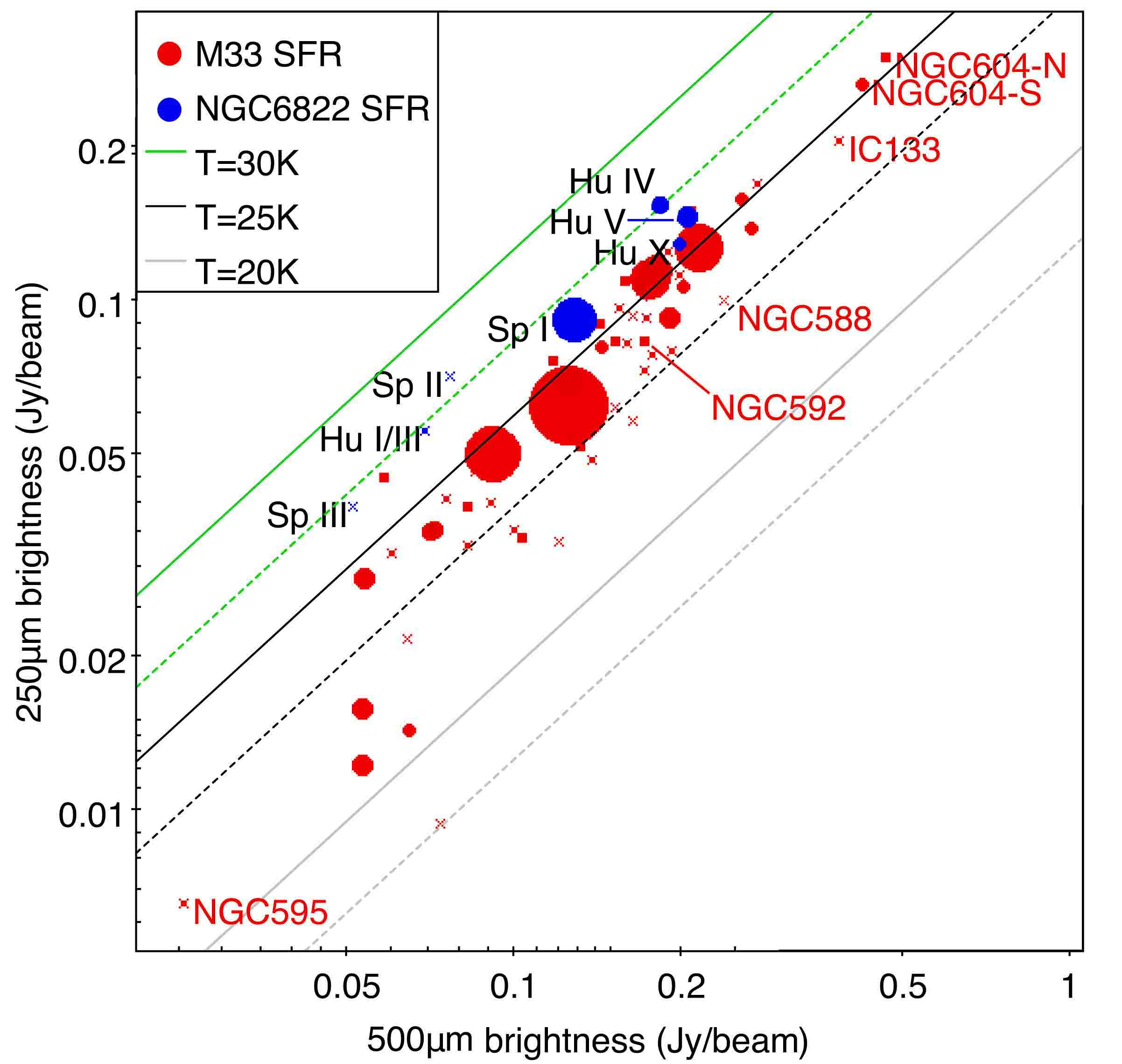}
    \includegraphics[width=7.1cm]{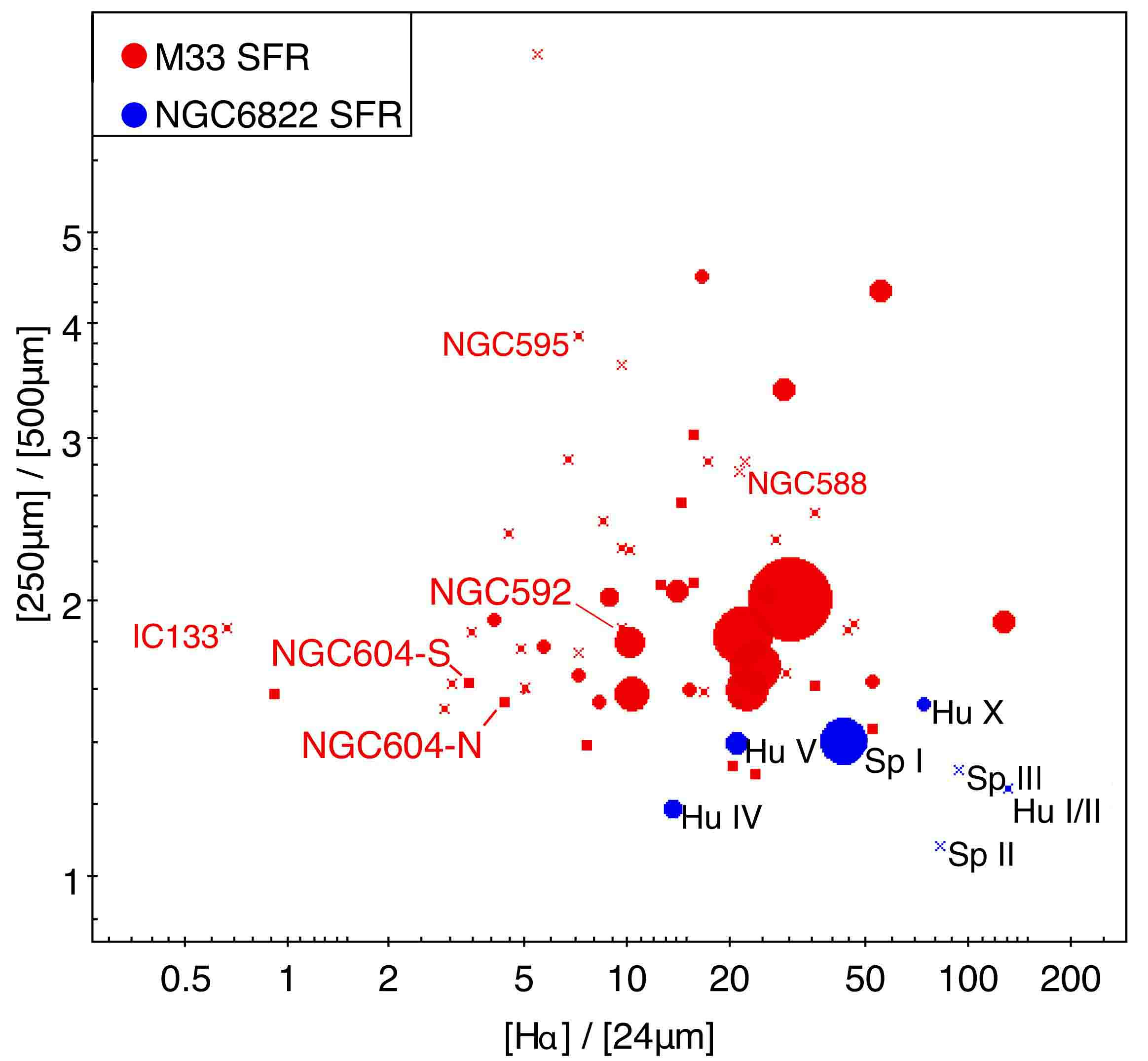}
    \caption{Photometric measurements for each SFR in M\,33 and NGC\,6822 (red and blue symbols respectively): H$\alpha$ and 24$\mu$m (upper panel), 250 and 500$\mu$m (middle), [H$\alpha$]$/$[24$\mu$m] and [250$\mu$m]$/$[500$\mu$m] (lower). The symbol size is proportional to the number of YSOs in each region (crosses mark particularly small regions); SFR radii for M\,33 and NGC\,6822 are respectively from our analysis and from \citet{2021MNRAS.507.5106K}. In the middle panel loci for modified blackbodies of different temperatures (colour-coded) and $\beta$\,=\,2 and 1.5 (solid and dashed lines respectively) are shown. Significant SFRs are labelled (see text).}
    \label{fig:ClusterMeasures}
\end{figure}

Figure\,\ref{fig:ClusterMeasures} shows SFR measurements in M\,33 and NGC\,6822: H$\alpha$ brightness against 24-$\mu$m brightness (upper panel), the far-IR 250- and 500-$\mu$m brightnesses (middle panel) and [H$\alpha$]$/$[24$\mu$m] against [250$\mu$m]$/$[500$\mu$m] ratios (lower panel). The H$\alpha$ and 24-$\mu$m brightnesses appear loosely correlated while the 250-$\mu$m and 500-$\mu$m brightnesses show a much tighter correlation (for M\,33 SFRs, $r_{\rm pearson}$\,$\sim$\,0.24 and 0.95 respectively). The 24-$\mu$m brightnesses for the SFRs in the two galaxies appear broadly consistent; the H$\alpha$ brightnesses for SFRs in NGC\,6822 are higher than those in M\,33, with none falling below $\sim$\,20\,counts per pixel. As noted in Sect.\,\ref{ssec:AncData}, the two H$\alpha$ images are taken with similar instruments and are calibrated in a consistent way \citep[see tables 1 and 2 of][]{2007AJ....134.2474M}, hence counts can be confidently compared between images. 

The higher H$\alpha$ brightnesses in NGC\,6822 may be a consequence of its lower metallicity \citep[$\sim$0.2\,Z$_\odot$, e.g.][]{1989MNRAS.240..563S,2007ApJ...658..328R}. At low metallicity, the interstellar medium (ISM) is more porous allowing for increased leakage of ionising radiation \citep{2006A&A...446..877M,2015A&A...580A.135D}. This effect has been used to explain the observed ISM properties in many dwarf galaxies \citep{2015A&A...578A..53C,2019A&A...626A..23C}. The resulting increased mean free path for far-UV photons could therefore make H$\alpha$-emitting regions in NGC\,6822 larger and brighter, compared to those in M\,33. 

In Fig.\,\ref{fig:ClusterMeasures} (middle panel) we show loci of theoretical modified blackbody emission, for dust temperatures 20, 25 and 30\,K (colour-coded) and values of $\beta$ the dust emissivity index ($\beta$\,=\,2 and 1.5, solid and dashed lines respectively). $\beta$ represents the frequency dependence of the dust emissivity which modifies the blackbody emission of dusty sources \citep{1983QJRAS..24..267H}. In NGC\,6822 values of $\beta$ adopted previously lie within this range \citep[e.g.][]{1996A&A...308..723I}. \citet{2014A&A...561A..95T} find that $\beta$ varies from $\beta$\,=\,2 in the central regions of M\,33 to $\beta$\,=\,1.3 in the outer disk; for SFRs however, a value of $\beta$\,=\,2 seems to be more appropriate \citep{2010A&A...518L..69B,2014A&A...561A..95T}. The position of the SFRs in NGC\,6822 is broadly consistent with those in M\,33, with a slight offset to higher 250-$\mu$m values. This offset corresponds to an increase in temperature of $\sim$\,2\,K or a variation in $\beta$ of $\sim$\,0.4. This offset could be due to the difference in dust properties, with ISM in NGC\,6822 having a smaller grain size than that in M\,33 \citep{2022arXiv220405548W}. Smaller grain sizes have been shown to correlate with higher grain equilibrium temperatures \citep{2020ApJ...904...38Z}. Dust temperatures have been found to be higher in the lower-metallicity SMC compared to LMC \citep{2010AJ....139.1553V}. Higher dust temperatures in dwarf galaxies can also lead to stronger far-IR emission per dust mass unit than in larger galaxies \citep{2022Galax..10...11H}. 

The symbol sizes in Fig.\,\ref{fig:ClusterMeasures} are proportional to the number of YSOs in the SFR; YSO numbers come from the {\sc DBSCAN} analysis in Sect.\,\ref{ssec:YSOclusters} for M\,33, and from table\,4 of \citet{2021MNRAS.507.5106K} for NGC\,6822 (these values are used instead of those reported by \citet{2019MNRAS.490..832J}, since PRF identification is also used). For the most populous regions in M\,33, measurements other than H$\alpha$ tend towards the ranges' averages (Fig.\,\ref{fig:ClusterMeasures} upper and middle panels). This could be expected if the largest SFRs identified by {\sc DBSCAN} are in fact comprised of multiple smaller regions of differing properties that even out when integrated. In the Milky Way the Orion-Eridanus superbubble contains several stellar subgroups, sites of ongoing star formation \citep[e.g.][]{2009AJ....137.3843B,2021PASJ...73S.239L} alongside structures with older populations \citep[e.g.][]{2008hsf1.book..459B}. As the individual subgroups evolve they expand into and interact with one another \citep{2015ApJ...808..111O}, creating large-scale substructures that have been mapped in free-streaming H$\alpha$ emission \citep{2015ApJ...808..111O,2022arXiv220500012H}. The Orion-Eridanus superbubble when scaled to the distance of M\,33 would be approximately 254\,pc (61\,arcsec) in size, which would place it well within the range of M\,33 SFRs (see Fig.\,\ref{fig:RadialPlots} and Appendix D). This may explain why the largest SFRs in M\,33 have the brightest H$\alpha$ emission but unremarkable overall mid- and far-IR brightness, appearing relatively evolved (see next section).

\subsubsection{SFR evolutionary status}
\label{sssec:SFRevol}

As previously mentioned we utilise SFRs in NGC\,6822 for which there is an established evolutionary sequence as a guide for the SFRs we identify in M\,33. Given the previously discussed differences between SFRs in M\,33 and NGC\,6822 and the very different sample sizes, we compared the SFRs in the two galaxies using the regions' rank order in each ratio. 

\begin{figure}
    \centering
    \includegraphics[width=8.35cm]{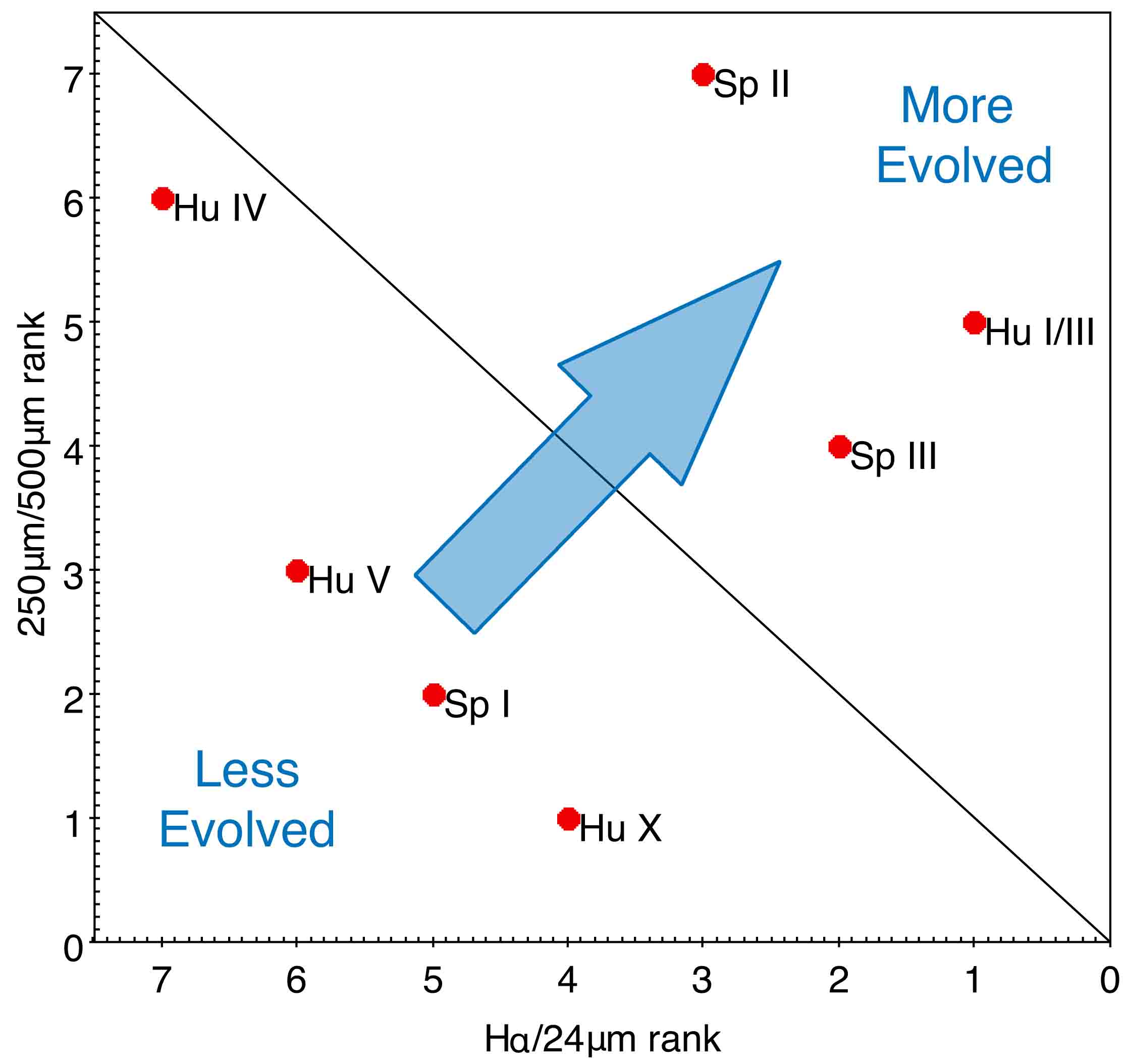}
    \includegraphics[width=\columnwidth]{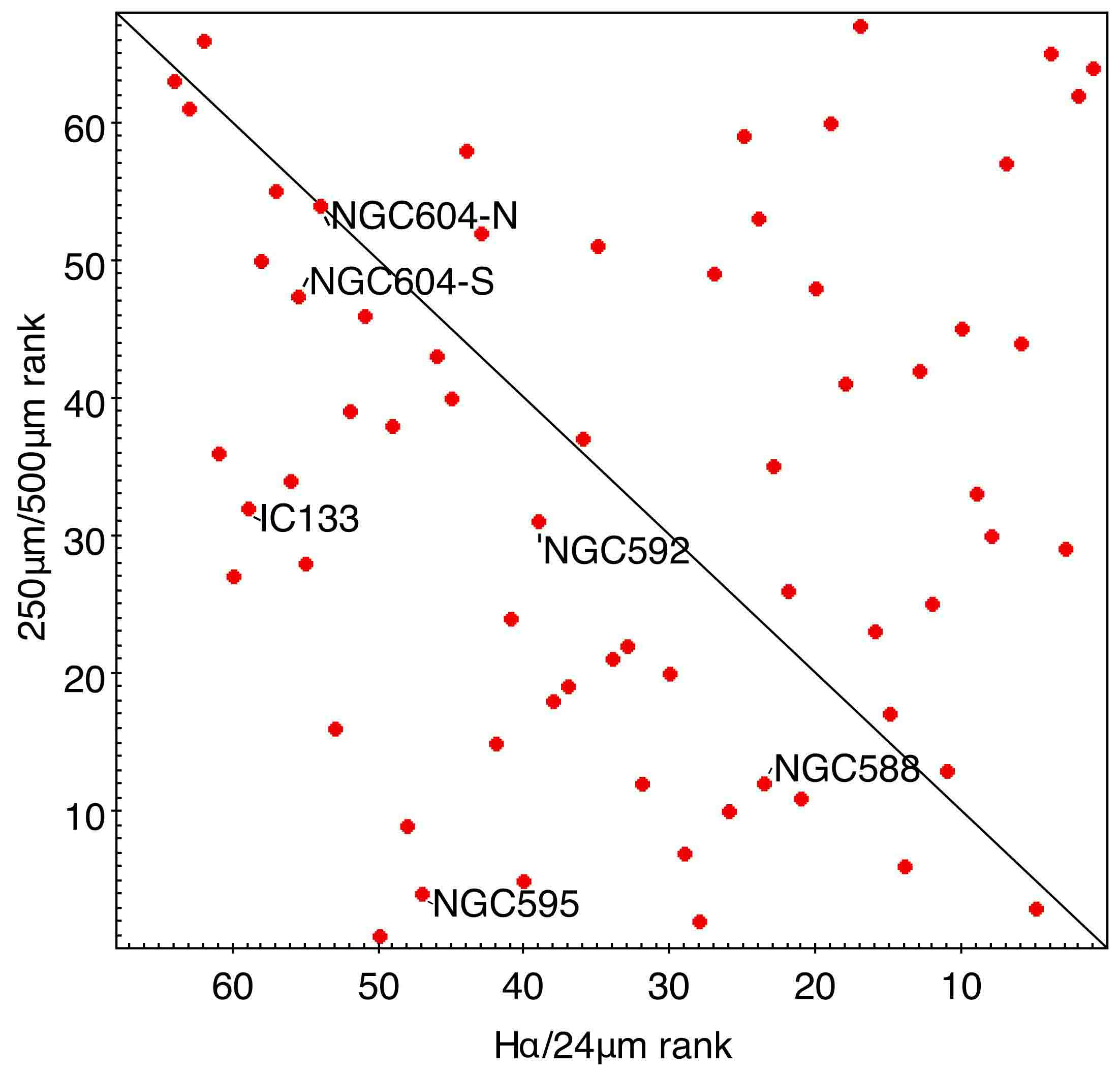}
    \caption{SFRs in NGC\,6822 (upper) and M\,33 (lower) shown by their relative ranks in the [H$\alpha$]$/$[24$\mu$m] and [250$\mu$m]$/$[500$\mu$m] ratios. The diagonal line indicates the locus of equal rank in both ratios. In the top panel the direction of SFR evolution is indicated by the arrow; significant SFRs are labelled (see text for more detail).}
    \label{fig:EvolRank}
\end{figure}

In Fig.\,\ref{fig:EvolRank} (upper panel) we show the rank sequence for the SFRs in NGC\,6822. Using a combination of [H$\alpha$]$/$[24$\mu$m] ratio and CO morphologies, \citet{2017ApJ...835..278S} suggest that {\it Hubble}\,I/III and {\it Hubble}\,X are likely more evolved than {\it Hubble}\,IV and {\it Hubble}\,V. \citet{2019MNRAS.490..832J} use similar tracers to propose that the most evolved SFR is likely {\it Hubble}\,I/III, {\it Spitzer}\,I and {\it Hubble}\,V are the least evolved and regions {\it Hubble}\,IV and X, {\it Spitzer}\,II and III are intermediate. This is broadly consistent with the position of the regions in Fig.\,\ref{fig:EvolRank}: the least evolved regions are found towards the lower left and most evolved towards the upper right; the blue arrow indicates the sequence of evolution. While this generally agrees with the relative evolution stages from \citet{2017ApJ...835..278S} and \citet{2019MNRAS.490..832J}, the exception is {\it Hubble}\,X which would appear less evolved in our analysis. Whilst the intermediate regions in NGC\,6822 appear quite distant from the locus of parity between the ranked ratios (shown by the black diagonal lines in Fig.\,\ref{fig:EvolRank}), this is due to the low number of SFR present. Indeed, this effect is not seen in the rank order of the SFRs in M\,33 (lower panel of Fig.\,\ref{fig:EvolRank}). Some of the most prominent H\,{\sc ii} regions and SFR in M\,33 are discussed further in Sect.\,\ref{sssec:IndvSFR}.

In order to compare the evolution stage of SFRs in M\,33 and NGC\,6822, we convert the distance from the locus of rank parity in Fig.\,\ref{fig:EvolRank} into a measure of evolution, normalised to the number of sources in each sample. We call this the evolution score. A negative evolution score represents a less evolved, more embedded region in which the [250$\mu$m]$/$[500$\mu$m] ratio dominates over the [H$\alpha$]/[24$\mu$m] ratio. A positive value of the normalised evolution score reflects a region in which the ISM is being cleared by bright young massive stars and neutral gas is ionised forming H\,{\sc ii} regions, allowing shorter-wavelength photons to freely propagate. 

\begin{figure}
    \centering
    \includegraphics[width=\columnwidth]{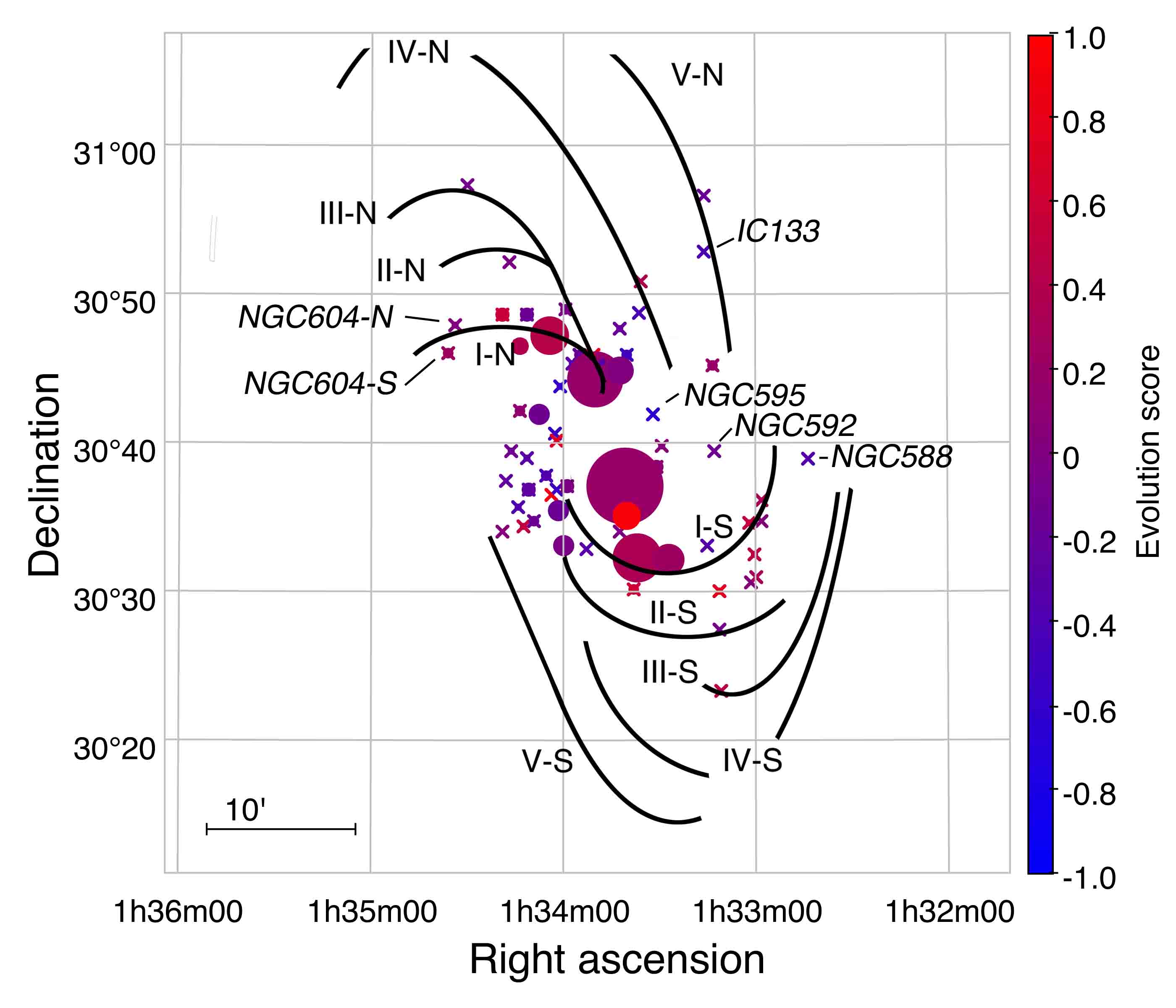}
    \caption{Galactic location of SFRs in M\,33 shown with a schematic labelled spiral structure. Symbol size is proportional to the number of YSOs, colour shows the evolution score (the smallest regions are marked with a cross). The least evolved regions (purple hues) ring the centre of the galaxy with more evolved regions (red hues) located further out in the disk (see also Fig.\,\ref{fig:RadEvol}). SFRs discussed in Sect.\,\ref{sssec:IndvSFR} are labelled.}
    \label{fig:SpiralEvol}
\end{figure}

\begin{figure*}
    \centering
    \includegraphics[width=16.5cm]{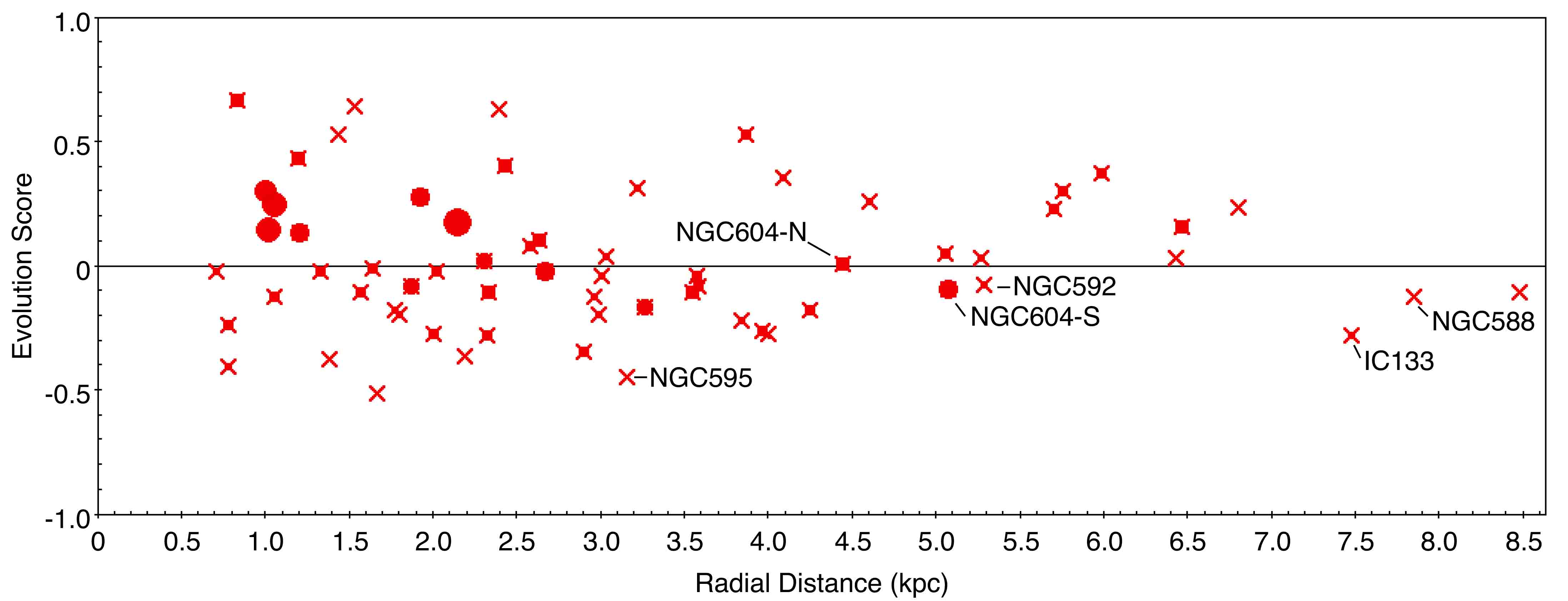}
    \caption{Normalised evolution score against radial distance for SFRs in M\,33. Symbol size is proportional to the radius of each cluster. IC\,133 and NGC\,588 are notable outliers in that they have a low evolution score and lie far out in the disk of M\,33.}
    \label{fig:RadEvol}
\end{figure*}

To characterise star formation activity across the disk of M\,33 we investigate the relation between galactic location and evolution score. In Fig.\,\ref{fig:SpiralEvol} the location of each SFR in M\,33 is shown superposed on spiral arm structure; region size and evolution score are indicated by symbol size and colour respectively. The largest regions, that are also generally the most evolved, lie at the base of the two primary spiral arms I-N and I-S; the least evolved SFRs mainly lie immediately surrounding the central region of the galaxy. In Fig.\,\ref{fig:RadEvol} we explore in more detail the effect of radial distance on the evolution scores of the SFRs. At radii larger than $\sim$\,4.5\,kpc most SFRs have positive evolution scores (i.e. are more evolved); obvious outliers are IC\,133 in arm V-N and NGC\,588 which are discussed further in Sect.\,\ref{sssec:IndvSFR}. 

\begin{figure*}
    \centering
    \includegraphics[width=16.5cm]{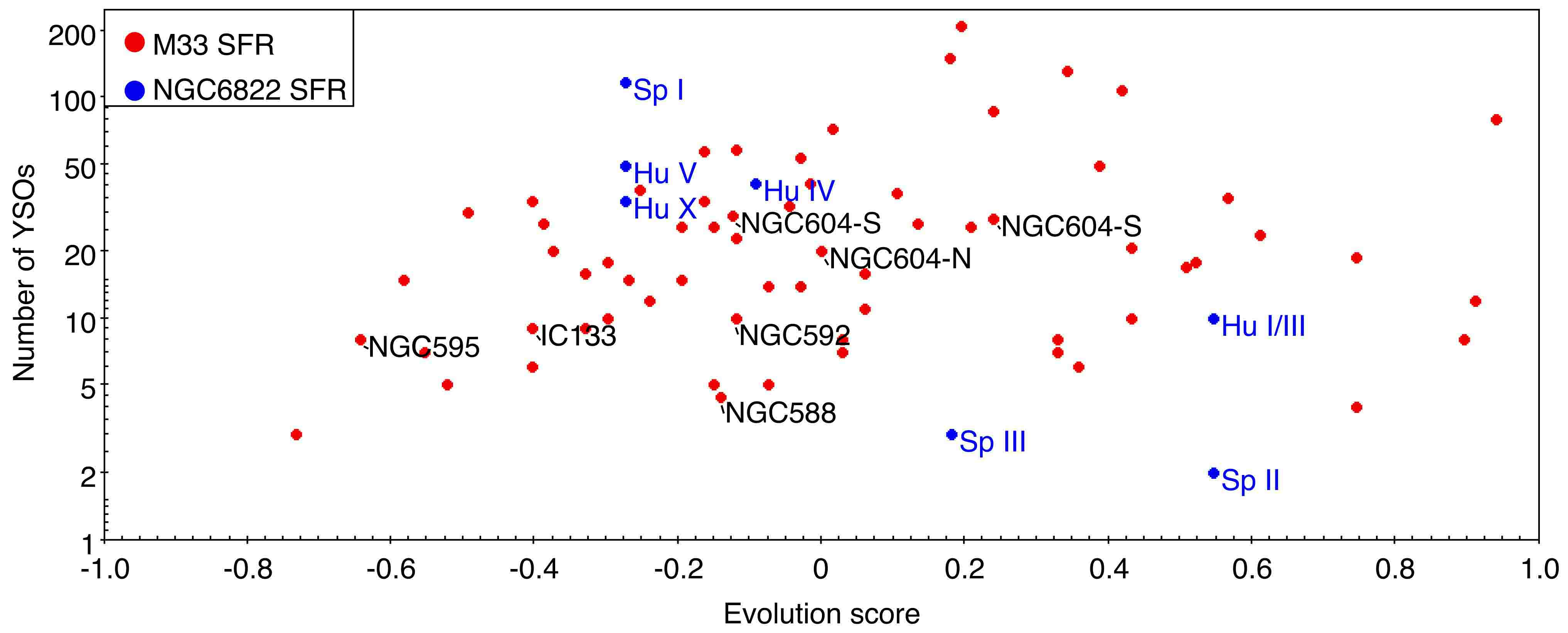}
    \caption{Number of YSOs against normalised evolution scores for SFRs in M\,33 and NGC\,6822. The number of YSOs for SFRs in M\,33 and NGC\,6822 are respectively from our analysis and from \citet{2021MNRAS.507.5106K}. There seems to be a slight tendency for larger SFRs to appear more evolved in M\,33.}
    \label{fig:ClstEvol}
\end{figure*}

We compare the relation between the number of YSOs in a SFR to its evolution score in both M\,33 and NGC\,6822 in Fig.\,\ref{fig:ClstEvol}. The SFRs in NGC\,6822 show a decreasing number of YSOs with increasing evolution score ($r_{\rm pearson}$\,$\sim$\,$-$0.71). For M\,33 the opposite trend is seen, albeit less strong ($r_{\rm pearson}$\,$\sim$\,0.21), that could suggest that larger regions appear more evolved. In order to assess the similarity of the two SFR samples we used a 2-Dimensional KS test \citep{1983MNRAS.202..615P,1987MNRAS.225..155F}. We find a low probability ($p$\,$\sim$\,0.29) that the two samples are drawn from distinct parent samples, with the caveat that the low number of SFRs analysed in NGC\,6822 is not an effect of sampling, since these are all the significant SFRs in this galaxy. Whilst the [H$\alpha$]$/$[24$\mu$m] ratio (lower panel of Fig.\,\ref{fig:ClusterMeasures}) suggests that larger regions should correlate to higher evolution scores, this is in fact not seen in Fig.\,\ref{fig:ClstEvol}, with the exception of the very largest regions ($n_{\rm YSOs}$\,$\geqslant$\,50); as discussed in Sect.\,\ref{sssec:SFObProp} these regions likely result from the combination of multiple smaller SFRs.

\subsubsection{SFRs in the context of GMCs}
\label{sssec:GMC}

We checked the positions of the 68 SFRs identified in our analysis against existing giant molecular cloud (GMC) catalogues. \citet{2017A&A...601A.146C} identified 566 GMCs using CO\,(2--1) observations and classify these according to their emission characteristics: the types A, B and C correspond respectively to inactive GMCs, clouds with embedded or low-mass star formation and clouds with massive or exposed star formation, the latter associated with H$\alpha$ and 24-$\mu$m emission. We find 17 type A, 16 type B, and 54 type C GMCs that have a positional overlap with 62 out of 68 SFRs ($\sim$\,91\,per\,cent), using the SFR median radii provided in Table \ref{tab:SFR} and the GMC deconvolved effective  radii \citep[see table\,5 of][]{2017A&A...601A.146C}. Since significant 24\,$\mu$m emission \citep[strongly correlated with star formation, e.g.][]{2018MNRAS.479..297W} is required for a type C classification, most SFRs are indeed matched to this GMC type; furthermore as discussed in Sect.\,\ref{ssec:YSOprop} our analysis allows only for the identification of the most massive YSOs. Type A matches occur mostly for the largest SFRs that in fact include multiple GMCs of different types. \citet{2017A&A...601A.146C} find that type-B GMCs are rarely found close to the spiral arms of M\,33, whereas types A and C are more closely aligned to H\,{\sc i} filaments in the arms. We do not find an overall correlation between GMC type and SFR evolution score.

Star formation in the two primary spiral arms of M\,33 has been previously studied to differing degrees. Arm I-N contains several well studied GMCs along its extension as well as the prominent H\,{\sc ii} region NGC\,604. We find counterparts to GMCs also identified in the CO\,(3--2) observations of M\,33 by \citet{2012ApJ...761...37M}: SFRs\,11 and 36 (GMC\,16 and 8 respectively in their nomenclature) as well as two additional CO peaks in between these (see figure\,1 of \citealt{2021ApJ...912...66K}), which correspond to SFRs 25 and 35. NGC\,604 is recovered in our analysis as two SFRs discussed further in Sect.\,\ref{sssec:IndvSFR}. These regions in I-N were studied in detail with recent ALMA observations \citep{2020ApJ...896...36T,2020ApJ...903...94M,2021ApJ...912...66K}. All the SFRs in arm I-N are associated with Type-C GMCs, SFR\,36 also contains a Type-B GMC. SFR\,11/GMC\,16 contains filamentary structure \citep{2020ApJ...896...36T}, which is not present in the comparatively inactive SFR\,36/GMC\,8 \citep{2021ApJ...912...66K}. The lack of filamentary structure, and the presence of a Type-B GMC in SFR\,36/GMC\,8 would suggest it is less evolved compared to SFR\,11/GMC\,16, as supported by the evolution scores, $-$0.11 and $-$0.03 respectively. 

Arm I-S is less disturbed than arm I-N and it seems to exhibit a clear progression from Type-A to Type-C GMCs through the arm \citep{2017A&A...601A.146C}. Due to the few SFR matches to Type-A GMCs we cannot confirm this observation. The progression across arm I-S, as well as spatial offsets between filamentary structures and H\,{\sc i} gas (e.g., in SFR\,11/GMC\,16, \citealt{2020ApJ...896...36T}), are consistent with the ``quasi-stationary spiral structure'' model of \citet{1964ApJ...140..646L}. Whilst \citet{2021ApJ...912...66K} find H\,{\sc i} gas velocities in SFR\,36/GMC\,8 which are consistent with ``dynamic spiral'' theory \citep{2014PASA...31...35D}, they cannot rule out an external source for the gas such as tidal interactions with M\,31 \citep{2018PASJ...70S..52T}. 

\subsubsection{Comments on individual M\,33 SFRs}
\label{sssec:IndvSFR}

NGC\,604 is one of the largest and brightest H\,{\sc ii} regions in the Local Group \citep[e.g.][]{2002MNRAS.329..481B}. Located around 4.8\,kpc from the centre of M\,33 in arm I-N, star formation has been studied there at many wavelengths \citep[e.g.][]{1983HiA.....6..611H,2012AJ....143...43F,2012ApJ...761...37M,2018PASJ...70S..52T,2020Galax...8...13L,2020ApJ...903...94M}. NGC\,604 has undergone multiple star formation events \citep{2011MNRAS.411..235E}, with earlier star formation episodes suggested to trigger the subsequent episodes \citep{2007ApJ...664L..27T,2018PASJ...70S..52T}.

\begin{figure*}
    \centering
    \includegraphics[width=16.5cm]{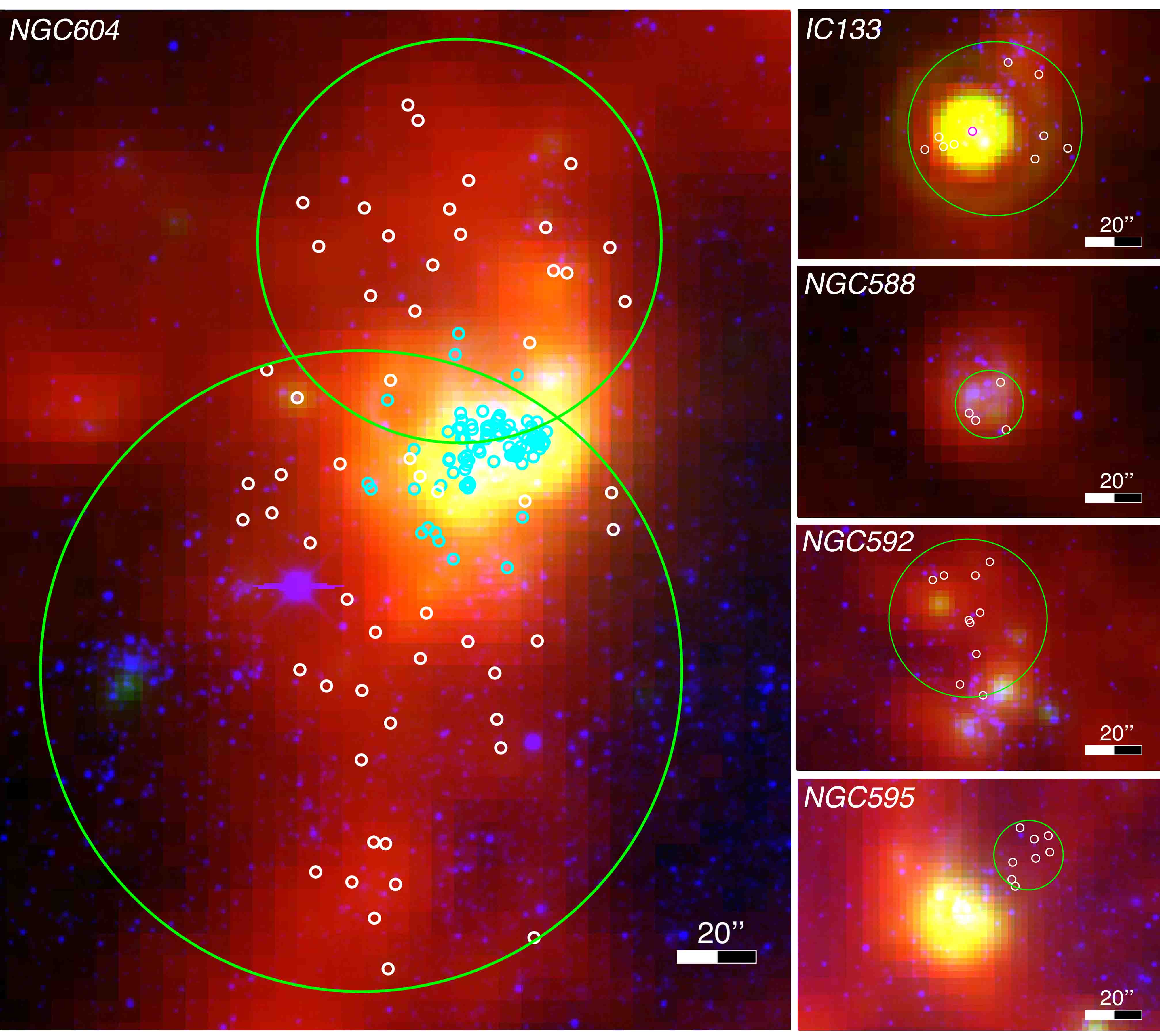}
    \caption{RGB image (250$\mu$m {\it Herschel}-SPIRE, 24$\mu$m {\it Spitzer}-MIPS, H$\alpha$ respectively $-$ see Sect.\,\ref{ssec:AncData} for image details) of NGC\,604, IC\,133, NGC\,588, NGC\,592 and NGC\,595. YSOs identified in this work are shown by white circles, the extent of each SFR is shown by the green circles, in NGC\,604 cyan circles show YSOs identified in \citet{2012AJ....143...43F}, in IC\,133 the magenta circle shows the location of the maser counterpart (see text).}
    \label{fig:NotedSFR}
\end{figure*}

Using GEMINI-NIRI photometry with excellent seeing conditions ($\sim$\,0.35\,arcsec), \citet{2012AJ....143...43F} identified 68 massive YSOs in the central region of NGC\,604 (see left panel of Fig.\,\ref{fig:NotedSFR}). Whilst all the YSOs identified by \citet{2012AJ....143...43F} are brighter than the catalogue sensitivity limits (see Sect.\,\ref{ssec:NIRData}), none of these sources have a counterpart within 1\,arcsec in the near-IR catalogue of \citet{2015MNRAS.447.3973J}. In fact within 30\,arcsec of the centre of NGC\,604 \citep[01:34:32.1, +30:47:01;][]{2015AJ....149...57M}, the near-IR catalogue contains only 27 sources, of which five are classified by the PRF analysis (as WRs, consistent with the young nature of the region). Likewise the {\it Spitzer}-IRS pointings described in \citet{2012ApJ...761....3M} are all located in this region of sparse near-IR point sources. This is a limitation of the catalogue used in our analysis in this region of extremely bright ambient emission; the YSOs we identify in our analysis are found instead at its periphery.

The {\sc DBSCAN} analysis divides NGC\,604 into two SFRs, North and South of the centre of brightest emission (see Fig.\,\ref{fig:NotedSFR}). The two SFRs (3 and 17 in Table\,\ref{tab:SFR}), which we refer to as NGC\,604-N and -S, contain 20 and 28 YSOs respectively. Whilst the separation of NGC\,604 into two SFRs may be in part driven by the paucity of near-IR data described above, this separation is supported astrophysically by the decomposition of NGC\,604 into multiple components in CO\,(1--0) and (2--1) emission \citep{2014A&A...567A.118D,2020ApJ...903...94M} and the South-East and North-West CO lobes of \citet{1992ApJ...385..512W} which are coincident with our SFRs. We record different evolution scores respectively 0.01 and $-$0.09 for NGC\,604-N and -S, indicative of star formation propagating from North to South in agreement with the \citet{2007ApJ...664L..27T} and \citet{2020ApJ...903...94M} scenarios of triggered star formation in NGC\,604. We note however that our analysis probes larger scales and in fact NGC\,604-N lies outside the region discussed in those literature analyses. It is therefore more relevant to consider the larger scale H\,{\sc i} gas interactions discussed in \citet{2018PASJ...70S..52T}. They identified two components of H\,{\sc i} gas separated by $\sim$\,20\,km\,s$^{-1}$; NGC\,604-N is co-spatial with a peak in the redshifted component whilst NGC\,604-S is co-spatial with the blue-shifted component (see their figure\,11). The collision of these two large H\,{\sc i} gas components is suggested to have triggered the star forming activity and growth of NGC\,604 \citep{2018PASJ...70S..52T}; such a scenario has also been proposed for other regions in arm I-N, namely SFR\,11/GMC\,16 and SFR\,36/GMC\,8 \citep{2021ApJ...912...66K}. The origin of the infalling gas is not clear, however the presence of a H\,{\sc i} stream between M\,33 and M\,31 \citep{2008MNRAS.390L..24B,2012AJ....144...52L} due to a previous interaction between these two galaxies \citep{2018ApJ...864...34S} offers one possible explanation \citep{2018PASJ...70S..52T}.

NGC\,595 (SFR\,47), in which we identify eight YSOs, is the second most luminous H\,{\sc ii} region in M\,33 after NGC\,604 \citep{2009ApJ...699.1125R} and is comparatively understudied. It lies to the North-West of the centre of M\,33 towards the base of arm IV-N. Its evolution score of $-$0.4 suggests that NGC\,595 is yet to reach peak star formation and may be amongst the youngest sites of star formation in the galaxy. The YSOs are located North-West of the bright 24-$\mu$m and 250-$\mu$m emission (see Fig.\,\ref{fig:NotedSFR}).

As noted in Sect.\,\ref{sssec:SFRevol} and Fig.\,\ref{fig:RadEvol}, the H\,{\sc ii} region IC\,133 (SFR\,62) has a low evolution score ($-$0.28) for its large radial distance ($\sim$7.5\,kpc). IC\,133 is located in arm V-N and contains nine YSOs, and a source of H$_2$O maser \citep{1988A&A...200...26H,1993ApJ...406..482G} and OH maser \citep{1987MNRAS.226..689S} emission. We identify a bright ($K_s$\,$=$\,14.4\,mag) and red ($J-K_s$\,$=$\,1.5\,mag) source as the likely near-IR counterpart of the maser emission (at a distance of $\sim$\,0$\rlap{.}^{\prime\prime}28$) at coordinates 01:33:16.54,\,$+$30:52:49.7 which the PRF classifies into several classes across the 100 runs: $n_{\rm RSG}$\,=\,74, $n_{\rm CAGB}$\,=\,18, $n_{\rm YSO}$\,=\,5, $n_{\rm AGN}$\,=\,3. This suggests that a RSG classification is the most likely, since such sources are also known to harbour water maser emission \citep[e.g.][]{1998A&A...337..141V}. The presence of an RSG source in an SFR is more likely if the H\,{\sc ii} region is more mature (>10\,Myr), such that stars can evolve sufficiently to become RSGs, which is not reflected by the evolution score for IC\,133. This may indicate that star formation in IC\,133 is restarting after a period of hiatus. 

Directly West of the centre of M\,33 and not obviously linked with any spiral arm, is the prominent H\,{\sc ii} region NGC\,592. This is SFR\,18 that contains ten YSOs. This H\,{\sc ii} region is thought to be young, with age estimates from far-UV SED fitting of 4 and 5.6\,Myr \citep[respectively][]{2006AJ....131..849P,2009MNRAS.394.1847U}. \citet{2009ApJ...699.1125R} find compact knots of H$\alpha$ coincident with the brightest 24\,$\mu$m sources. We assign NGC\,592 an evolution score of $-$0.08. 

NGC\,588, another large H\,{\sc ii} region in which star formation has been studied \citep[e.g.][]{2009ApJ...699.1125R,2011MNRAS.413.2242M} lies almost directly West of NGC\,592, between the tips of arms I-S and III/IV-S as indicated in Fig.\,\ref{fig:SpiralEvol}. Only 4 YSOs are classified within its extent (SFR\,68). Alongside IC\,133, NGC\,588 is notable for its low evolution score ($-$0.12) at high radial distance ($\sim$7.8\,kpc) from the centre of M\,33 (see Fig.\,\ref{fig:RadEvol}).

\subsection{YSO masses and star formation rate}
\label{ssec:YSOprop}

The properties of the YSO sources analysed here are likely dominated by the most massive source in an unresolved proto-cluster \citep[see also discussions in][]{2013MNRAS.428.3001O, 2016MNRAS.455.2345W, 2017MNRAS.464.1512W}. This effect on YSO model fitting analysis is discussed in \citet{Chen_2010}, and accordingly \citet{2019MNRAS.490..832J} present their mass estimates for YSOs in NGC\,6822 as overestimated for the dominant source but underestimated for the total unresolved cluster. Furthermore it is also widely accepted that most massive stars are found in binaries or multiple systems \citep[e.g.][]{2008MNRAS.386..447S,2012Sci...337..444S,2014ApJS..213...34K}, implying that the dominant source is in turn an unresolved binary. These important caveats affect similar analysis in the literature (e.g. \citealt{2013ApJ...778...15S,2019MNRAS.490..832J} respectively in the SMC and NGC\,6822) and are impossible to account for properly, and thus the mass estimates we discuss below should be taken with some caution.

Since the YSOs identified in our analysis only have photometry in the three near-IR bands, it is not feasible to obtain their masses using individual SED fitting as seen in, e.g., \citet{2008AJ....136...18W,2013ApJ...778...15S,2019MNRAS.490..832J}. We therefore use predicted near-IR $K_s$-band magnitudes (scaled to the distance of M\,33) and $J-K$ colours estimated from the model grid of \citet{2006ApJS..167..256R} and the YSOs' positions in the CMD to assign them a model mass. For each of the 4985 YSOs identified by the PRF we thus obtained a mass estimate as described below. Due to the depth of the near-IR catalogue (see Sect.\,\ref{ssec:NIRData}) our analysis is likely sensitive to only the most massive YSOs. Given these sources evolve rapidly onto the main sequence once they leave their embedded stages, we use only models in the grid corresponding to Stage\,0/I YSOs. We note that this model grid does not represent a realistic mass distribution in an Initial Mass Function (IMF) sense.

Each YSO is compared to models within a 0.5\,mag distance in CMD space. For YSOs with at least three models in this range the median mass for the models is adopted; for YSOs with fewer models within 0.5\,mag distance the closest three models are used to compute the median model mass. This latter group of YSOs accounts for $\sim$\,11\,per\,cent of all YSOs and $\sim$\,10\,per\,cent of YSOs assigned to clusters; we consider these mass estimates more uncertain. YSO mass estimates range from 6\,$-$\,27\,M$_\odot$ with a median value of 13\,M$_\odot$. 

\begin{figure}
    \centering
    \includegraphics[width=\columnwidth]{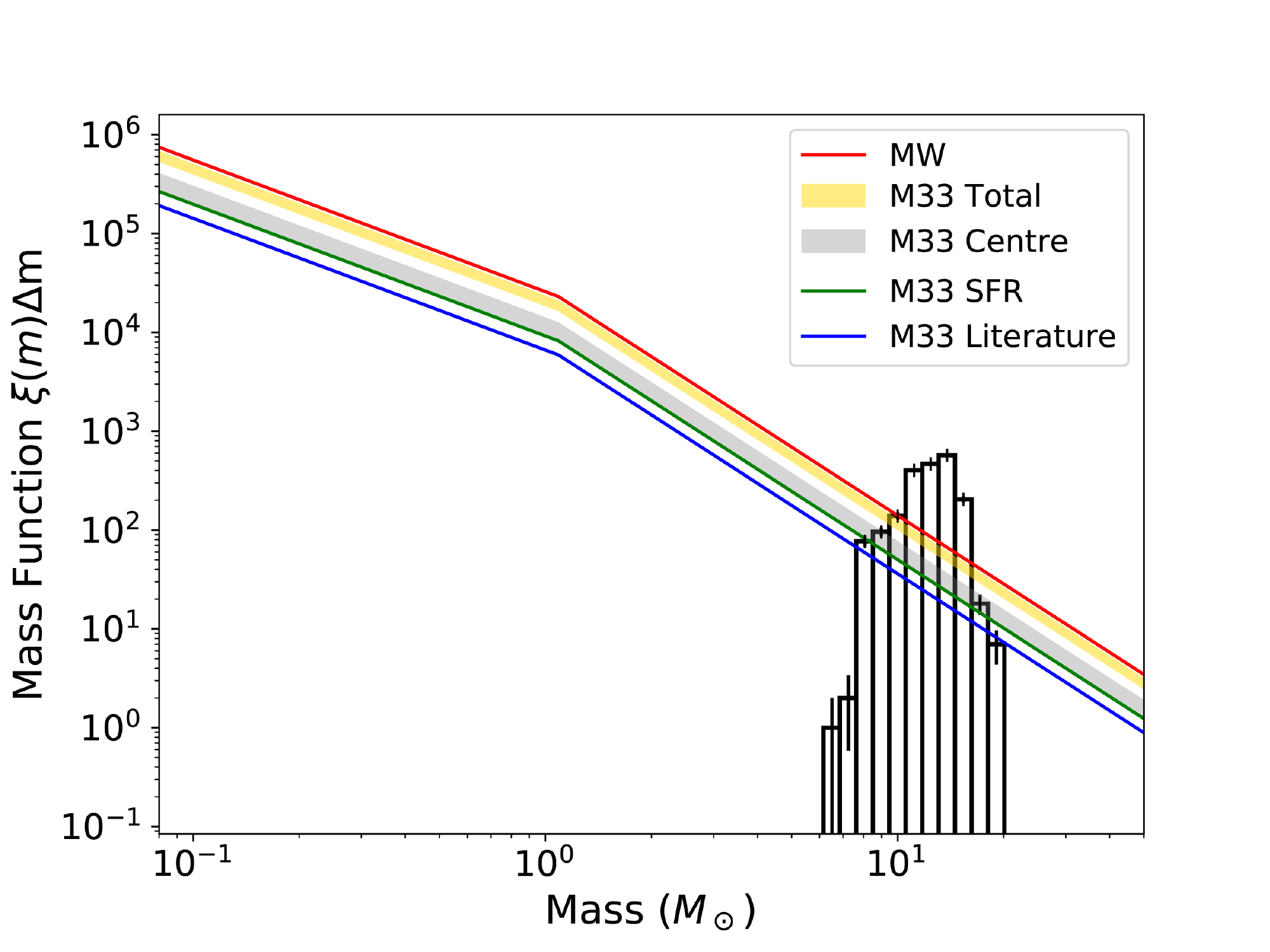}
    \caption{The mass distribution of the 1986 YSOs assigned to M\,33 SFRs, with scaled \citet{2002Sci...295...82K} IMFs overlain, see text for details. Poisson errors are indicated for each histogram bin.}
    \label{fig:IMF}
\end{figure}

The mass distribution of the YSOs assigned to SFRs is shown in Fig.\,\ref{fig:IMF}, with a total mass of $2.5\times10^4$\,M$_\odot$. Using the commonly adopted functional form for the IMF by \citet{2002Sci...295...82K}, scaled to match the observed mass distribution, and integrated over the range 0.08\,$-$\,100\,M$_\odot$, we estimate the total mass of YSOs in SFRs as $1.5\times10^5$\,M$_\odot$. Adopting a Stage 0/I lifetime of 0.2\,Myr \citep[e.g.][and references therein]{2019MNRAS.490..832J} we estimate a star formation rate of 0.63\,M$_\odot$\,yr$^{-1}$ in M\,33's SFRs (green line in Fig.\,\ref{fig:IMF}). Due to the effects of crowding, the lower PRF classification certainty, and potential contamination (see Sect.\,\ref{ssec:YSOclusters}), we estimate the star formation rate separately for the unclustered YSOs in the central region. This rate is 0.79\,$\pm$\,0.16\,M$_\odot$\,yr$^{-1}$ (grey shaded region in Fig.\,\ref{fig:IMF}). Considering all YSOs, the total star formation rate is 1.42\,$\pm$\,0.16\,M$_\odot$\,yr$^{-1}$ (gold shaded region), that overlaps with Milky\,Way estimates.

There are numerous determinations of global star formation rates in the Milky Way (MW), as compiled in table\,1 of \citet{2011AJ....142..197C} for a range of methods (ionisation rates, supernovae rates, near-IR to far-IR dust-heating ratios, nucleosynthesis rates and YSO counts), re-scaled to a \citet{2002Sci...295...82K} IMF; typical values are in the range $\sim$\,1.9\,$\pm$\,0.4\,M$_\odot$\,yr$^{-1}$ \citep[see also][]{2018ApJS..237...33X}. More recent work that uses Bayesian statistics to compare the rates compiled by \citet{2011AJ....142..197C} favours a rate of 1.65\,$\pm$\,0.19\,M$_\odot$\,yr$^{-1}$ as the best fit to the data \citep{2015ApJ...806...96L}. Using direct YSO counts, \citet{2011MNRAS.416..972D} find a rate of 1.75\,$\pm$\,0.25\,M$_\odot$\,yr$^{-1}$ (the average shown as the red line in Fig.\,\ref{fig:IMF}). The rate of star formation in star forming galaxies is strongly correlated to the mass of available gas \citep{2012ARA&A..50..531K}. It is therefore expected that M\,33 ($M_{\rm gas}$\,$\sim$\,3\,$\times$\,10$^{9}\,$M$_\odot$,\,\citealt{2003MNRAS.342..199C}) has a lower star formation rate than the MW ($M_{\rm gas}$\,$\sim$\,5\,$\times$\,10$^{10}$\,M$_\odot$,\,\citealt{2015ApJ...806...96L}) as seen in Fig.\,\ref{fig:IMF}.

Star formation rates estimated from direct YSO counts tend to be higher than those calculated with other methods that are sensitive to different star formation timescales, as documented in the MCs \citep[e.g.][]{2010ApJ...721.1206C,2012A&A...542A..66C} and NGC\,6822 \citep{2019MNRAS.490..832J}, but are generally consistent \citep{2013ApJ...778...15S}. Our estimates are higher than the values calculated using the 24\,$\mu$m (0.2\,M$_\odot$\,yr$^{-1}$), H$\alpha$ (0.35\,M$_\odot$\,yr$^{-1}$) and far-UV (0.55\,M$_\odot$\,yr$^{-1}$) emission maps by \citet{2009A&A...493..453V}, that adopted an average value of 0.45\,M$_\odot$\,yr$^{-1}$. More recently far-UV {\it Hubble} Space Telescope observations of M\,33 were used by \citet{2022arXiv220611393L} to find a star formation rate of 0.74\,M$_\odot$\,yr$^{-1}$ over the last 100\,Myr. The Long-Period Variable (LPV) population gives an estimated star formation rate of 0.42\,M$_\odot$\,yr$^{-1}$ over the last 100\,Myr \citep{2017MNRAS.464.2103J}. \citet{2019MNRAS.483..931E} explored star formation in M\,33 at multiple scales from 49\,pc to 782\,pc at mid and far-IR wavelengths and estimated star formation rates of 0.44\,$\pm$\,0.1\,M$_\odot$\,yr$^{-1}$ (at 100\,$\mu$m) and 0.34{\raisebox{0.5ex}{\tiny$\substack{+0.42 \\ -0.27}$}}\,M$_\odot$\,yr$^{-1}$ (at 12\,$\mu$m). Using CO and HCN relations, \citet{2006ApJ...650..933B} inferred an integrated star formation rate in M\,33 of 0.7\,M$_\odot$\,yr$^{-1}$. Our estimates for the star formation rate of M\,33 are broadly consistent with these estimates, towards the upper end.

\section{Conclusions}
\label{sec:Conc}

In this work, we identified and described the YSO population across the whole disk of the flocculent spiral galaxy M\,33 for the first time.
We adapted the PRF classification technique which was successfully applied in NGC\,6822 \citep{2021MNRAS.507.5106K} to better reflect the stellar populations in M\,33. The PRF classifier was trained using a combination of near-IR and far-IR feature information to identify nine target classes. 

In total we applied the PRF to 162,746 sources of which 66,378 are consistently assigned to the same class across a total of 100 PRF runs. The PRF classifies with a median estimated accuracy of 86\,per\,cent (the accuracy is based on the PRF's confusion matrices for the test runs). A total of 4985 YSOs were identified. A {\sc DBSCAN} clustering analysis of the YSO population was used to identify 68 SFRs, mostly previously unknown, across the disk of M\,33, containing 1986 YSOs. Most of these SFRs are located in the spiral arms. 2437 YSOs are found in the central $\sim$\,11.6\,$\times$\,10.4\,arcmin$^2$ region, that is too crowded for the clustering algorithm to be effective. The remainder 562 YSOs are seemingly isolated based on our analysis.

In total 62 out of our 68 SFRs ($\sim$\,91\,per\,cent) are co-spatial with GMCs identified by \citet{2017A&A...601A.146C}, mainly Type-C clouds ($\sim$\,87\,per\,cent) with tracers of massive or exposed star formation. We identify SFR counterparts to the prominent H\,{\sc ii} regions IC\,133, NGC\,588, NGC\,592, NGC\,595 and NGC\,604. A novel approach combining [H$\alpha$]$/$[24$\mu$m] and [250$\mu$m]$/$[500$\mu$m] ratios was used to constrain the comparative evolutionary status of the M\,33 SFRs, using regions in NGC\,6822 as a benchmark sample. These ratios was converted into a common metric for ease of comparison. This evolution score was used to compare SFRs in the context of radial distance in the galaxy, number of YSOs and the relation to M\,33's spiral structure. 

We resolve the wider NGC\,604 environment into two SFRs with different evolutionary status; these are co-spatial with two different H\,{\sc i} gas components identified by \citet{2018PASJ...70S..52T}. The collision of these components may explain the triggering of initial star formation and progression from North to South \citep{2007ApJ...664L..27T}, for which we see some evidence in our evolution score analysis. In this scenario the in-falling H\,{\sc i} gas is responsible for feeding the growth of NGC\,604 into one of the most luminous H\,{\sc ii} regions in the Local Group. This gas component may originate from a stream connecting M\,33 and M\,31 arising from an earlier interaction with M\,31. 

We used model grids for Stage\,0/I YSOs \citep{2006ApJS..167..256R} to estimate the mass of each of the 4985 YSOs. Given that a SED fitting analysis is not feasible with just three near-IR bands, masses are derived from the models that are closest to each YSO in the colour-magnitude diagram. Estimated YSO masses range from 6\,$-$\,27\,M$_\odot$ with a median value of 13\,M$_\odot$. The total mass of YSOs assigned to SFRs is $2.5\times10^4$\,M$_\odot$. Using a Stage\,0/I lifetime of 0.2\,Myr, we estimate a star formation rate of 0.63\,M$_\odot$\,yr$^{-1}$ for M\,33 spiral arms' SFRs. In the central region of M\,33 we find a higher value of 0.79\,$\pm$\,0.16\,M$_\odot$\,yr$^{-1}$ with the caveat of less certain source classifications for this crowded region. These estimates give a total M\,33 star formation rate of 1.42\,$\pm$\,0.16\,M$_\odot$\,yr$^{-1}$ determined from direct YSO counts. As expected from gas mass scaling relations, the star formation rate for M\,33 is lower than that of the more massive MW (1.75\,$\pm$\,0.25\,M$_\odot$\,yr$^{-1}$, \citealt{2011MNRAS.416..972D}, also computed from YSO counts).

We have for the first time identified massive YSOs on galactic scales in a Local Group spiral galaxy, extending such analysis beyond the nearby star-forming dwarf galaxies (LMC, SMC and NGC\,6822). Machine learning approaches, as we have demonstrated, offer an invaluable tool for disentangling and classifying large data sets. The next generation of observatories such as the Extremely Large Telescope, {\it James Webb} and {\it Roman Space Telescopes} will deliver a treasure-trove of such data, extending the range of galaxies in which such studies can be conducted.

\section*{Acknowledgements}

The authors would like to thank the anonymous referee for their helpful comments and suggestions that helped improve the paper. DAK acknowledges financial support from STFC via their PhD studentship programmes as well as Keele University via their Phase-2 COVID-19 funding extension programme.
The authors thank A. Javadi for help with gaining access to their catalogues.
{\it Herschel} is a European Space Agency (ESA) space observatory with science instruments provided by European-led Principal Investigator consortia and with important participation from NASA.
This work is based on observations made with the {\it Spitzer Space Telescope}, which is operated by the Jet Propulsion Laboratory, California Institute of Technology under a contract with NASA. 
This work has made use of data from the ESA mission
{\it Gaia} (\url{https://www.cosmos.esa.int/gaia}), processed by the {\it Gaia}
Data Processing and Analysis Consortium (DPAC,
\url{https://www.cosmos.esa.int/web/gaia/dpac/consortium}). Funding for the DPAC
has been provided by national institutions, in particular the institutions
participating in the {\it Gaia} Multilateral Agreement.

\section{Data Availability}
The data underlying this article which are not included in the supplementary online materials will be shared on reasonable request to the corresponding author.



\bibliographystyle{mnras}
\bibliography{Refs} 




\appendix

\section{On-line materials}
\begin{enumerate}
    \item Down-sampling of the largest training set classes
    \item PRF classification statistics
    \item De-projected coordinates 
    \item Additional SFR statistics
    \item Central region spatial distributions
\end{enumerate}


\bsp	
\label{lastpage}
\end{document}